\documentclass[10pt,a4paper]{scrartcl}

\usepackage{CLICdp}
\usepackage{CLICdp_definitions}
\usepackage{float}
\usepackage[disable]{todonotes} 
\usepackage{adjustbox}
\usepackage{sansmath}
\usepackage{multicol}
\usepackage{pgf-pie}
\usepackage{pgfplots, pgfplotstable}
\usetikzlibrary{positioning}
\usepackage{threeparttable}
\usepackage{multirow}
\usepackage{makecell}
\usepackage{etoolbox}
\usepackage[section]{placeins}
\usepackage{afterpage}
\usepackage{siunitx} 
\robustify\bfseries
\usepackage{wrapfig}
\newcommand{\imgc}[1]{(image credit: #1)}
\newcommand{\imcl}{\imgc{CLIC}}

\title{The Compact Linear e$^+$e$^-$ Collider (CLIC)}

\notitlestamp 

\date{\today}

\onbehalfof{ {\em Input to the European Strategy for Particle Physics Update 2025} \\ \vspace*{0cm} {\em on behalf of the CLIC and CLICdp Collaborations} \\ \vspace*{.4cm} {\large 26 May 2025} }  

\addauthor{Contact persons: E.\,Adli,\thanks{Erik.Adli@cern.ch}, D.\,Dannheim, A.\,Robson, S.\,Stapnes \vspace{.5cm}}{\\}

\addauthor{Editors: E.\,Adli, G.\,D'Auria, N.\,Catalan Lasheras, V.\,Cilento, R.\,Corsini, D.\,Dannheim, S.\,Doebert, M.\,Draper, A.\,Faus-Golfe, E.\,MacTavish, A.\,Grudiev, A.\,Latina, L.\,Linssen, J.\,Osborne, Y.\,Papaphilippou, P.\,Roloff, A.\,Robson, C.\,Rossi, A.\,Sailer, D.\,Schulte, E.\,Sicking, S.\,Stapnes, I.\,Syratchev, R.\,Tomas Garcia, W.\,Wuensch}{\\}

\DeclareSIUnit\years{years}
\DeclareSIUnit\days{days}

\abstract{The Compact Linear Collider (CLIC) is a TeV-scale high-luminosity linear e$^+$e$^-$ collider studied by the international CLIC and CLICdp collaborations hosted by CERN. CLIC uses a two-beam acceleration scheme, in which normal-conducting high-gradient 12\,GHz accelerating structures are powered
via a high-current drive beam.
For an optimal exploitation of its physics potential, CLIC is foreseen to be built and operated in stages. The initial stage will have a centre-of-mass energy
of 380\,GeV, with a site length of 11 km. The 380\,GeV stage optimally combines the exploration of Higgs and top-quark physics, including a top threshold scan near 350\,GeV. A higher-energy stage, still using the initial single drive-beam complex, can be optimised for any energy up to 2 TeV. Parameters are presented in detail for a 1.5 TeV stage, with a site length of 29 km.
Since the 2018 ESPPU reporting, significant effort was invested in CLIC accelerator optimisation, technology developments and system tests, including collaboration with and gaining experience from new-generation light sources and free-electron lasers. CLIC implementation aspects at CERN have covered detailed studies of  civil engineering, electrical networks, cooling and ventilation, scheduling, and costing. 
The CLIC baseline at 380\,GeV is now 100 Hz operation, with a luminosity of 4.5$\times 10^{34}$\,cm$^{-2}$s$^{-1}$ and a power consumption of 166\,MW. Compared to the 2018 design, this gives three times higher luminosity-per-power. The new baseline has two beam-delivery systems, allowing for two detectors operating in parallel, sharing the luminosity.  
The cost estimate of the 380\,GeV  baseline is approximately 7.2 billion CHF.
The construction of the first CLIC energy stage could start as early as $\sim$2034-2035 and beam commissioning and first beams would follow a decade later,
marking the beginning of a physics programme spanning 20-30 years and providing
excellent sensitivity to Beyond Standard Model physics, through direct searches and via a
broad set of precision measurements of Standard Model processes, particularly in the Higgs and top-quark sectors. This report summarises the CLIC project, its implementation and running scenarios, with emphasis on new developments and recent progress. 
It concludes with an update on the CLIC detector studies and on the physics potential in light of the improved accelerator performance. The physics potential includes results from the 3 TeV energy stage, which was studied in detail for the CLIC CDR in 2012 and the CLIC Project Implementation Plan of 2018.}

\graphicspath{ {./logos/}{./figures/} }

\usepackage{cleveref}
\newlength{\abc}
\AtBeginDocument{%
\renewcommand{\ref}[1]{\mbox{\Cref{#1}}}

}

\begin{document}

\titlepage

\section{Introduction}
\label{sec:introduction}

The Compact Linear Collider (CLIC) is a mature and detailed proposal for a multi-\si{\TeV} high-luminosity linear e$^+$e$^-$ collider, developed by the CLIC accelerator collaboration~\cite{clic-study}.
CLIC uses a novel two-beam acceleration technique, with normal-conducting accelerating structures
operating in the range of \SIrange{72}{100}{\mega\volt/\meter}.
Detailed studies of the physics potential and detector for CLIC, and R\&D on detector technologies, have been carried out by the CLIC detector and physics (CLICdp) collaboration~\cite{clicdp}. 

The CLIC Conceptual Design Report (CDR) was published in \num{2012}~\cite{cdrvol1,cdrvol2,cdrvol3}. 
The main focus of the CDR was to demonstrate the feasibility of the CLIC accelerator at \SI{3}{\TeV},
and to confirm that high-precision physics measurements can be performed 
in the presence of particles from beam-induced backgrounds.  Following the completion of the CDR, 
detailed studies on Higgs and top-quark physics, with particular focus on the first energy stage,
concluded that the optimal centre-of-mass energy for the first stage of CLIC is $\roots=\SI{380}{\GeV}$. The 2018 ESPP Update~\cite{ESU18PiP, ESU18Summary} presented the accelerator design for a staged scenario, with operation at \SI{380}{\GeV}, \SI{1.5}{\TeV}, and \SI{3}{\TeV}~\cite{StagingBaseline}.  

Since 2018, important technical progress related to X-band technology and klystron design has been achieved. Integrated beam simulations have concluded on a 50\% increase in luminosity at \SI{380}{\GeV}, with respect to previous evaluations. New power estimates for the \SI{380}{\GeV} and \SI{1.5}{\TeV} collider result in significant energy savings, allowing CLIC to operate at a factor of two increased bunch-train repetition rate of 100 Hz, at a power of 166 MW at \SI{380}{\GeV}. Compared to the 2018 values, this represents a total factor of three increase in luminosity at equal power for the initial stage.

\noindent \begin{minipage}{\linewidth}
  \begin{minipage}{0.46\textwidth}
\noindent 

An interaction region with two beam delivery systems hosting two detectors has been designed. 
Taking the \SI{380}{\GeV} and \SI{1.5}{\TeV} stages as an example, CLIC proposes a scenario with a running period of 10 years each, which results in the integrated luminosities listed in~\ref{tab:clicstaging} for the two experiments combined, including luminosity ramp-up time at each stage.  CLIC provides $\pm 80$\% longitudinal electron polarisation and proposes a sharing between the two polarisation states at each energy stage for optimal physics reach~\cite{Roloff:2645352}.
  
  \end{minipage}
  \hspace*{.75cm}
  \begin{minipage}{0.46\textwidth}
    \vspace*{-1.25cm}
    \begin{table}[H] \centering
      \caption{Baseline CLIC energy stages and integrated luminosities, $\mathcal{L}_{\textrm{int}}$, for the initial stage and a high-energy stage at 1.5 TeV~\cite{RobsonNoteHiggs25}. \label{tab:clicstaging}}
      \begin{tabular}{SSS}\toprule
        {Stage} & {$\sqrt{s}$ [\si{\TeV}]} & {$\mathcal{L}_{\textrm{int}}$ [\si{\per\ab}]} \\
        \hline
        1 &  0.38{ (and 0.35)} &  4.3 \\
        2 &  1.5             &  4.0 \\
        \bottomrule
      \end{tabular}
    \end{table}
  \end{minipage}
\end{minipage}

This document summarises the current progress of the CLIC studies. 
There have been many updates in accelerator design, technology development, and system tests,
described in detail in~\cite{Adli:ESU25RDR}.
Further experience has been gained from the expanding field of Free Electron Laser (FEL) linacs and new-generation light sources. 
The CLIC design parameters are well understood and achieved in beam tests, confirming that the performance goals are realistic.
The implementation of CLIC at CERN has been updated and detailed since its 2018 description~\cite{ESU18PiP, ESU18Summary}, 
including civil engineering, electrical networks, cooling and ventilation, and scheduling. Systematic studies have put emphasis on optimising cost and energy efficiency.

A related strategy submission, the Linear Collider Facility@CERN~\cite{LCFCERNdraft}, discusses options for a future linear e$^+$e$^-$ collider, using various acceleration technologies that are under development. 
These options are all compatible with the CLIC underground civil engineering footprint presented here.


\ref{sec:accelerator} provides an overview of the CLIC accelerator design and performance
at \SI{380}{\GeV}, describes the path to higher energies, and gives an overview of the developments. 
\ref{sec:accelerator} also outlines key achievements from beam experiments and hardware tests,
providing evidence that the performance goals can be met. 
The present plans for the CLIC implementation, with emphasis on the \SI{380}{\GeV} stage, are given, 
as well as estimates of the energy consumption and of the cost for construction and operation. Further details on the accelerator design and performance are given in~\cite{Adli:ESU25RDR}.
In~\ref{sec:detector} the CLIC detector performance requirements and the CLIC detector design are described~\cite{CLICdet_note_2017, CLICdet_performance}, with reference to relevant advances in detector technology development. \Cref{sec:physics} presents a summary of the CLIC physics potential in the context of the updated CLIC accelerator performance. A summary and outlook are presented in~\ref{sec:summary}.




\section{CLIC Accelerator}
\label{sec:accelerator}

\begin{figure}[ht]
\begin{center}
\includegraphics[width=0.9\textwidth]{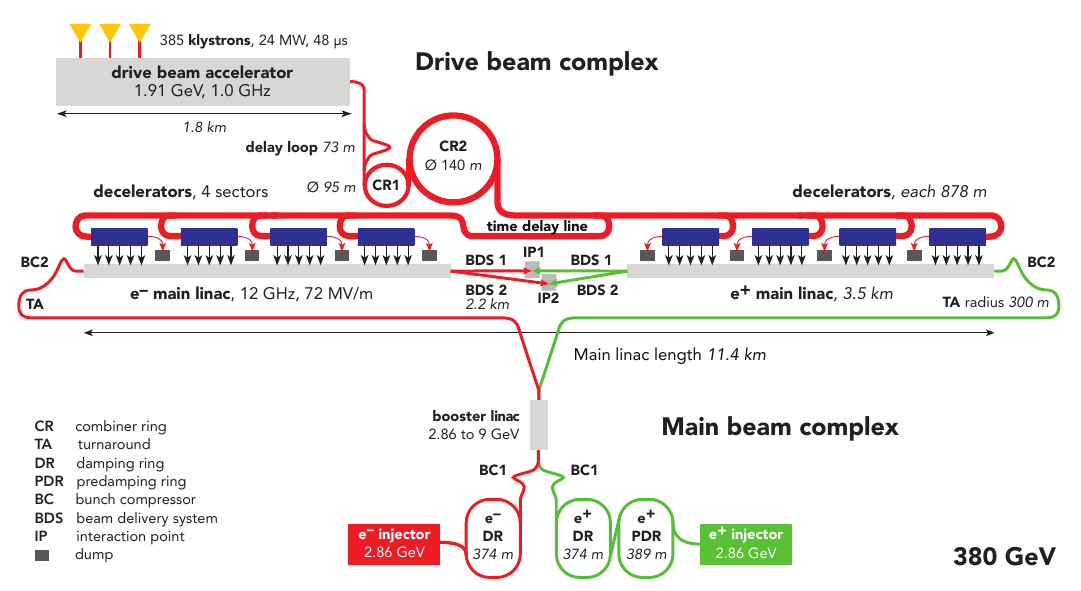}
\caption{Schematic layout of the CLIC complex at \SI{380}{\GeV}, with a double beam-delivery system operating with two detectors. \imcl}
\label{fig:clic_layout}
\end{center}
\end{figure}

The design luminosity at 380\,GeV is $\mathcal{L}=\SI{4.5e34}{\per\centi\meter\squared\per\second}$; a factor of three improvement with respect to 2018~\cite{ESU18PiP}.  The increase comes from reduced power consumption, and subsequent doubling of the repetition rate to 100 Hz; as well as updated studies on the luminosity performance, detailed below. A double beam delivery system that allows two interaction regions has been designed. The 2025 baseline configuration is to operate with two detectors, sharing the luminosity between them. The CLIC detector concept is compatible with the 2025 updated baseline.
A schematic overview of the accelerator configuration for the initial 380 GeV stage is shown in~\ref{fig:clic_layout}.

A design for a high-energy stage at 1.5 TeV, running at 50 Hz with a single drive beam, has also been worked out in detail. Parameters for the inital stage and the high-energy stage are given in~\ref{t:CLIC_stages_main}.

\begin{table}[!htb]
\caption{Key parameters for 380\,GeV and 1.5\,TeV stages of CLIC.  Parameters for energy options at 250\,GeV and 550\,GeV are also given; for these options the power and luminosity are scalings, based on the 380\,GeV and 1.5\,TeV designs.
\newline
$^*$The luminosity for the 1.5 TeV machine has not been updated to reflect recent alignment studies~\cite{PhysRevAccelBeams.23.101001}. If the same method is applied, the luminosity at 1.5\,TeV is expected to reach $\SI{5.6e34}{\per\centi\meter\squared\per\second}$.
}
\label{t:CLIC_stages_main}
\centering
\begin{tabular}{llllll}
\toprule 
Parameter  & Unit  & \textbf{380\,GeV}  &\textbf{1.5\,TeV}  & 250\,GeV  & 550\,GeV \tabularnewline
\midrule 
Centre-of-mass energy  & \si{\GeV}  & 380  & 1500  & 250  & 550 \tabularnewline
Repetition frequency  & \si{\Hz}  & 100  & 50 & 100  & 100 \tabularnewline
Nb. of bunches per train  &  & 352  & 312 & 352  & 352 \tabularnewline
Bunch separation  & \si{\ns}  & 0.5  & 0.5 & 0.5  & 0.5 \tabularnewline
Pulse length  & \si{\ns}  & 244  & 244 & 244  & 244 \tabularnewline
\midrule 
Accelerating gradient  & \si{\mega\volt/\meter}  & 72  & 72/100 & 72  & 72 \tabularnewline
\midrule 
Total luminosity  & \SI{e34}{\per\centi\meter\squared\per\second}  & 4.5  & 3.7$^{*}$  & $\sim$3.0  & $\sim$6.5 \tabularnewline
Lum. above \SI{99}{\percent} of $\sqrt{s}$  & \SI{e34}{\per\centi\meter\squared\per\second}  & 2.7  & 1.4  &  $\sim$2.1  &  $\sim$3.2 \tabularnewline
Total int. lum. per year  & fb$^{-1}$  & 540  & 444  & $\sim$350  & $\sim$780 \tabularnewline
Power consumption  & MW  & 166  & 287  & $\sim$130  & $\sim$210 \tabularnewline
\midrule 
Main linac tunnel length  & \si{\km}  & 11.4  & 29.0 & 11.4  & $\sim$15 \tabularnewline
Nb. of particles per bunch  & \num{e9}  & 5.2  & 3.7 & 5.2  & 5.2 \tabularnewline
Bunch length  & \si{\um}  & 70  & 44 & 70  & 70 \tabularnewline
IP beam size  & \si{\nm}  & 149/2.0  & 60/1.5 & $\sim$184/2.5  & $\sim$124/1.7 \tabularnewline
\bottomrule
\end{tabular}
\end{table}

\paragraph{Design and performance at 380\,GeV and extension to higher energies}

\begin{figure}[h]
\centering
\begin{subfigure}{.40\textwidth}
\subcaptionbox{\label{fig:IMP_1}}{\includegraphics[width=\textwidth]{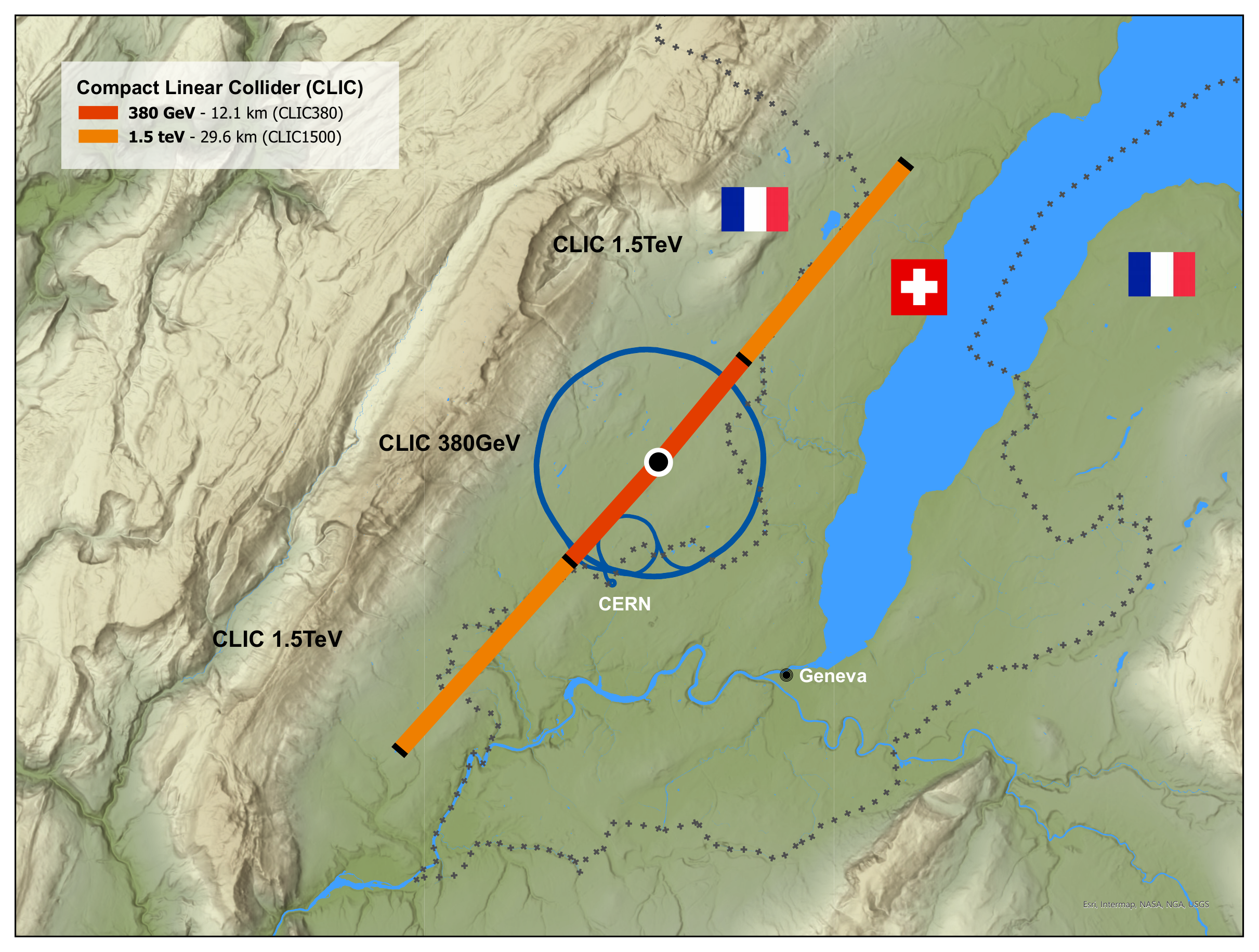}} 
\end{subfigure}
\hspace*{1cm}
\begin{subfigure}{.30\textwidth}
\subcaptionbox{\label{fig:CEIS_7a}}{\includegraphics[height=1.0\textwidth]{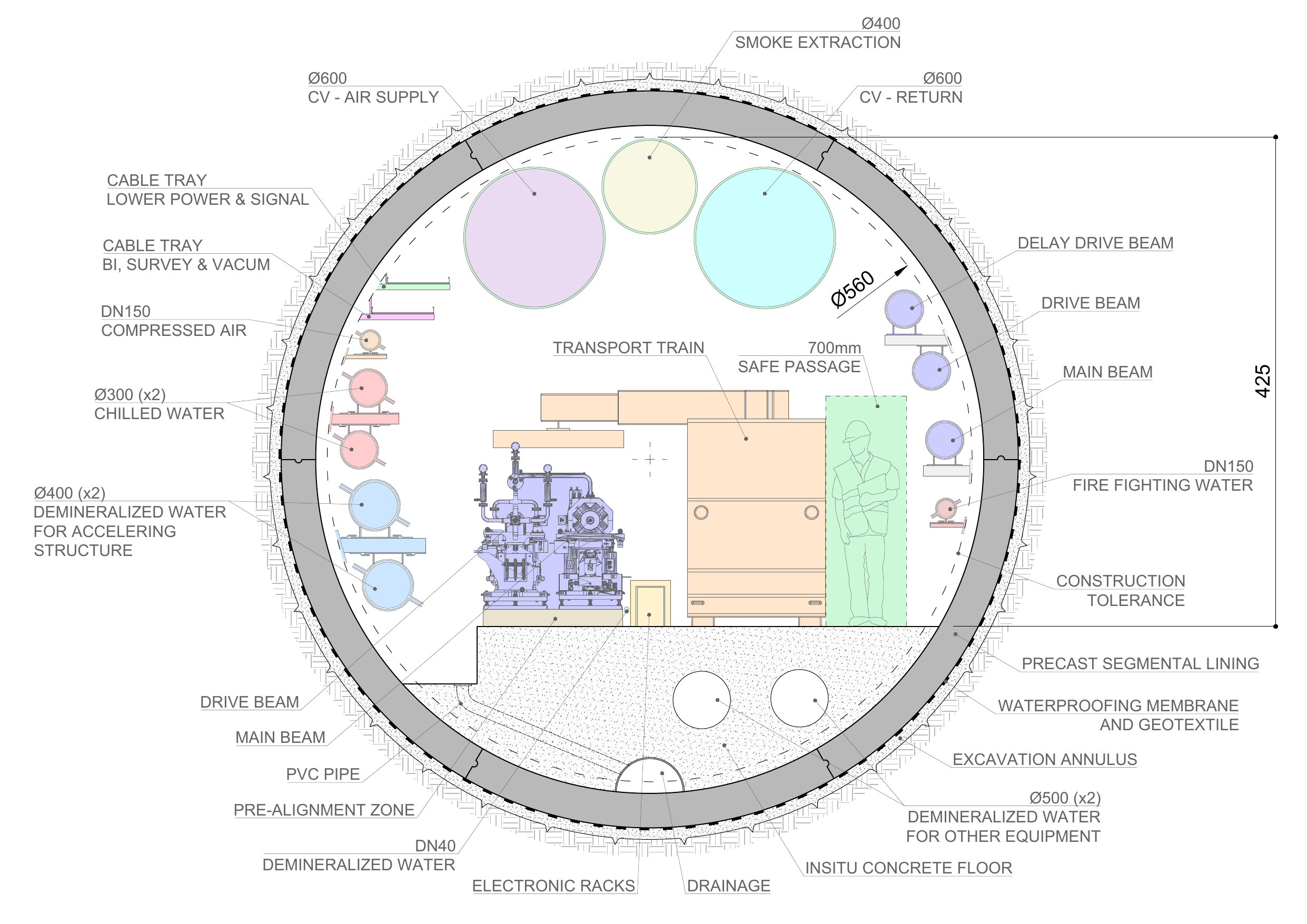}}
\end{subfigure}
\hspace*{1.5cm}
\caption{(a) The CLIC main linac footprint near CERN, showing the inital 380 GeV and a 1.5\,TeV high-energy stage. (b) The CLIC main linac tunnel cross section. The internal tunnel diameter is 5.6 m. \imcl}
\end{figure}

The location of the CLIC 380\,GeV machine has been optimised while still considering the requirements for a higher energy stage. This optimisation took into account the availability of existing CERN sites, and the regional geology and local environment. Previous experience from the construction of LEP and the LHC has shown that the sedimentary rock in the Geneva basin, known as molasse, provides suitable conditions for tunneling. Therefore, boundary conditions were established so as to avoid the karstic limestone of the Jura mountain range and to avoid siting the tunnels below Lake Geneva (see \ref{fig:IMP_1}) whilst maximising the portion of tunnel located in the molasse.

Based on the regional, geological and surface data, and using a decision-support web application for future accelerator localisation study (Geoprofiler), a 380\,GeV solution has been found that can be upgraded to the higher energy stage of 1.5\,TeV. 
The 380\,GeV stage is located entirely in molasse rock and avoids the complex Allondon depression. This solution is optimised for 380\,GeV and provides an excellent upgrade possibility to the 1.5\,TeV stage. A key advantage of this solution is that the interaction point and injection complex are located on the CERN Pr\'{e}vessin site.  With the exclusion of the 3 TeV stage, the Gland depression is no longer a geological concern. Thus there is scope at these lower energy stages of 380 GeV and 1.5 TeV to reduce shaft depths, still being housed in good Molasse rock. 

The main electron beam is produced in a conventional radio frequency (RF) injector, which allows polarisation. The beam emittance is then
reduced in a damping ring.
To create the positron beam, an electron beam is accelerated to 2.86 GeV and sent into a conventional tungsten target generating photons that produce electron--positron pairs. The positrons are captured and accelerated to 2.86\,GeV.
Their beam emittance is reduced, first in a pre-damping ring and then in a damping ring.
The ring to main linac system accelerates both beams to \SI{9}{\GeV}, compresses their bunch length, and delivers the beams to the main linacs. 
The main linacs accelerate the beams to the collision energy of \SI{190}{GeV}.
The beam delivery system removes transverse tails and off-energy particles with collimators, and compresses the beam to
the small size required at the interaction point (IP).
After collision, the beams are transported by the post-collision lines to their respective beam dumps.

To reach multi-TeV collision energies in an acceptable site length and at affordable cost, 
the main linacs use normal-conducting X-band accelerating structures;
these achieve a high accelerating gradient of \SI{100}{\mega\volt/\meter}.
For the first energy stage, a lower gradient of \SI{72}{\mega\volt/\meter} is the optimum to achieve the luminosity goal, which
requires a larger beam current than at higher energies.

In order to produce and support high gradients, the accelerating structures are required to be fed by short,
very high-power RF pulses, which are difficult to generate at 
acceptable cost and efficiency using conventional klystrons.
In order to provide the necessary high peak power, the novel drive-beam scheme uses low-frequency klystrons
to efficiently generate long RF pulses and to store their energy
in a long, high-current drive-beam pulse.
This beam pulse is used to generate many short, even higher intensity pulses that are distributed alongside the main linac,
where they release the stored energy in power extraction and transfer structures (PETS) 
in the form of short RF power pulses, transferred via waveguides into the accelerating structures. 
This concept strongly reduces the cost and power consumption compared with powering the structures directly by klystrons.

The upgrade to higher energies involves lengthening the main linacs:  connecting new tunnels to the existing tunnels, 
moving the existing modules to the beginning of the new tunnels, and adding new, higher-gradient modules. 
When upgrading to \SI{1.5}{\TeV}, the length of the beam delivery system (BDS) needs to be increased and new magnets installed.
The central main-beam production complex needs only minor modifications, to adjust to the smaller
number of bunches and smaller bunch charge.
The central drive-beam complex needs to be slightly extended to increase the drive-beam energy.
This staged collider can be implemented at CERN, as shown in~\ref{fig:IMP_1}.
The main-beam and drive-beam production facilities are located at the CERN Pr\'evessin site.
The main linac tunnel cross section is shown in~\ref{fig:CEIS_7a}.

\paragraph{Energy options}

A 250\,GeV option could be implemented using the 380\,GeV design as basis, removing 35\% of the two-beam modules in each sector, and reducing the drive beam energy correspondingly. The lower number of modules and the lower-energy drive beam linac give significant cost savings if 250 GeV were to be run as a first stage. The upgrade to 380 GeV would then be straight-forward; installing the missing modules, and extending the drive beam accelerator. 

With a single drive beam and running at 100 Hz, the 380\,GeV machine could be upgraded to 550 GeV.  
With a single drive beam and running at 50 Hz, the machine can be upgraded to approximately 2 TeV. 
A detailed description of a 3\,TeV stage is documented in the CLIC CDR~\cite{cdrvol1}. 

~\ref{t:CLIC_stages_main} also contains numbers for a 250 GeV and 550 GeV CLIC, for which the power and luminosity are scalings, based on the 380 GeV and 1.5 TeV designs. 

The CLIC beam energy can be adjusted to meet different physics requirements. In particular, during the 380\,GeV running, a period of operation around \SI{350}{GeV} is foreseen to scan the top-quark pair-production threshold.
Operation at much lower energies can also be considered. At the Z-pole, between \SI{7.5}{\per\fb} and \SI{135}{\per\fb} can be achieved per
year for an unmodified and a modified collider, respectively. The natural times to run at the highest Z-pole luminosity, which requires at modified layout, are the very beginning of the first stage, or during the transition to the second stage.


\paragraph{Operation with two detectors}

The 2025 CLIC baseline is to operate two detectors simultaneously, at an average repetition rate of 50\,Hz. This provides each detector with an average luminosity of about 2.2$\times 10^{34}$\,cm$^{-2}$s$^{-1}$. The doubling of machine total repetition rate from 50\,Hz to 100\,Hz, with respect to the previous baseline \cite{ESU18PiP} is achieved without major design changes, and with increases in the overall power consumption by about 60\%. The luminosity delivery can be flexible: if desired, 100\% of the luminosity can be provided to a single detector for any given period of time.

\begin{figure}[ht]
\centering
\includegraphics[width=0.7\textwidth]{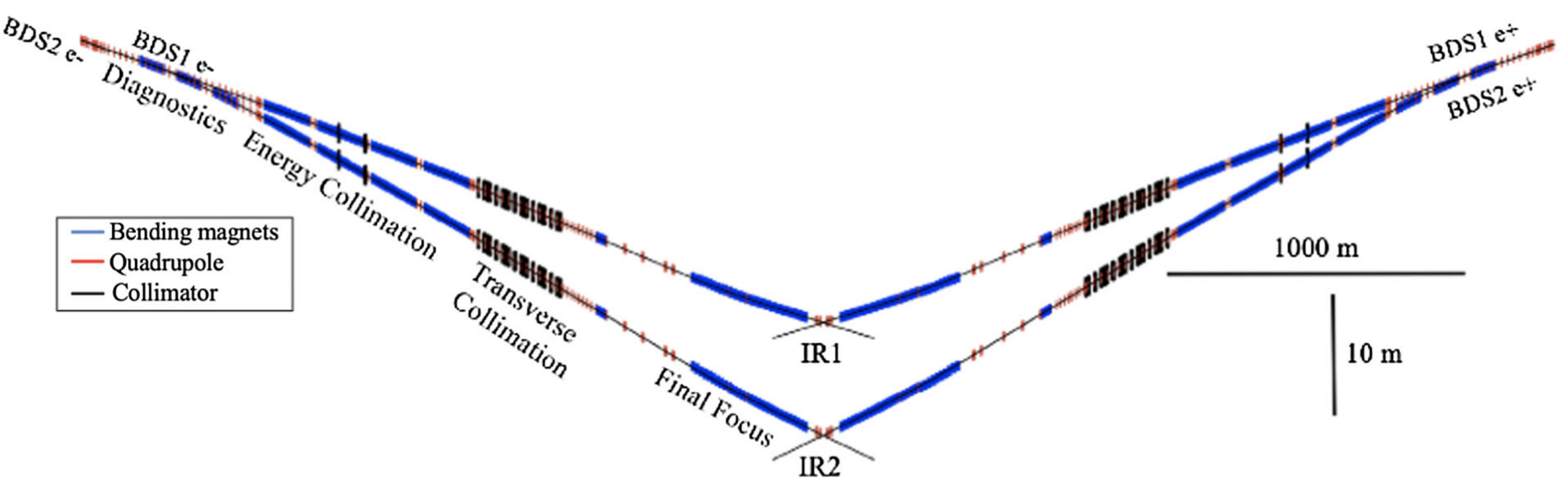}
\caption{Schematic layout of CLIC operating with two detectors. From~\cite{Cilento2021}.}. 
\label{scd:clic_2BDS_A}
\end{figure}

The operation of two detectors is achieved with the concept of a dual Beam Delivery System (BDS) serving two interaction regions (IRs) simultaneously~\cite{Cilento2021}.  The dual BDS introduces separate paths for the electron and positron beams to accommodate two detectors with distinct crossing angles. 
The 380 GeV stage of CLIC features a dual BDS design achieved by extending the diagnostics section of the baseline BDS. Eight additional FODO cells, each with a phase advance of 45$^{\circ}$, and with a total additional length of 300\,m, were added to separate the two IRs longitudinally and transversely. The two BDS paths (BDS1 and BDS2) are optimized for their respective IRs: the crossing angle of IR1 is 18\,mrad and the crossing angle of IR2 is 27\,mrad, which is also compatible with potential gamma-gamma collision configurations. A full schematic of the layout is shown in ~\ref{scd:clic_2BDS_A}.

To minimize synchrotron radiation effects, the bending angles and magnet strengths in the BDS have been carefully optimized. The transverse separation between IRs is approximately 10\,m, with a longitudinal offset of 40~m to ensure sufficient space for detector placement.

\paragraph{Updated power consumption}

The CLIC power consumption is very significantly reduced 
due to optimisation of the injectors and accelerating structures for 380\,GeV, a re-design of the damping ring RF systems, higher efficiency L-band klystrons as discussed in~\cite{Adli:ESU25RDR} and generally improving the RF efficiencies, and consistently using
the expected operational values instead of the full equipment capacity in the estimates. 

\begin{figure}[h!tb]
\centering
\includegraphics[width=0.7\textwidth]{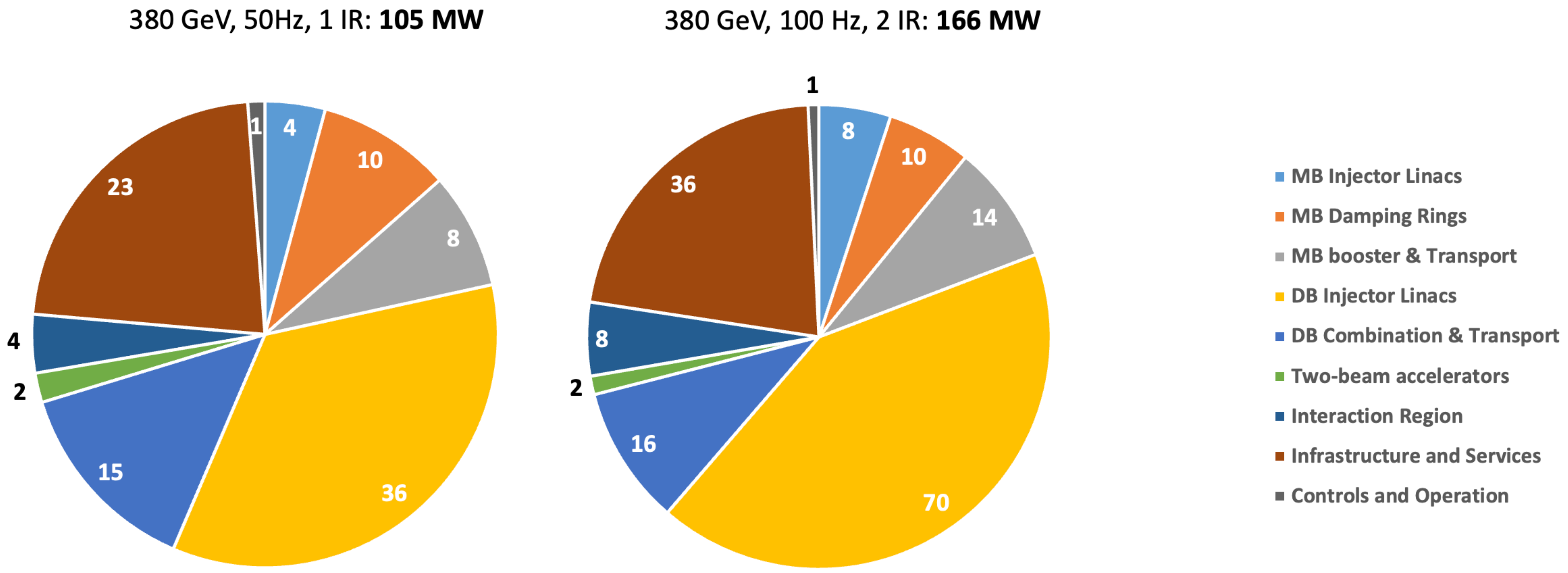} 
\caption{\label{fig_IMP_1} Breakdown of power consumption between different domains of the CLIC accelerator in MW, at a centre-of-mass energy of 380\,GeV. For the 100 Hz baseline machine, the contributions add up to a total of 166\,MW. For a 50 Hz machine, the contributions would add up to 105 MW.}
\end{figure}

The nominal power consumption has been estimated based on the 
detailed CLIC work breakdown structure. This yields, for the 380 GeV 100 Hz baseline option, with two interaction regions, a total of 166\,MW for all accelerator systems and services, taking into account network losses for transformation and distribution on site.  For the 50 Hz reduced power machine, the total is 105\,MW, assuming a single interaction region.  The breakdown per domain in the CLIC complex (including experimental area and detector) and per technical system is shown in ~\ref{fig_IMP_1}. Most of the power is used in the drive-beam and main-beam injector complexes; comparatively little in the main linacs. Among the technical systems, the RF represents the major consumer.  The reductions have also been applied to the 1.5\,TeV stage, resulting in a power consumption of 287\,MW, see details in~\cite{Adli:ESU25RDR}.

\paragraph{Luminosity performance}
The baseline plan for operating CLIC includes a yearly shutdown of 120 days, 30 days of commissioning,
20 days for machine development, and 10 days for planned technical stops.
This leaves 185 days of operation for collider physics data-taking. Assuming an availability during normal running of 75\%, this results in an integrated
luminosity per year equivalent to operating at full luminosity for \SI{1.2e7}{\second}~\cite{Bordry:2018gri}.   This leads to the integrated luminosities given in~\ref{tab:clicstaging}.

In order to achieve high luminosity, CLIC requires very small beam sizes at the collision point, as listed in~\ref{t:CLIC_stages_main}.
The resulting high charge density leads to
strong beam--beam effects, which result in the emission of beamstrahlung and production of background particles. This is limited to an acceptable level by
using flat beams, which are much larger in the horizontal than in the vertical plane. The key to high luminosity lies in the
small vertical beam size; therefore a small vertical emittance and strong vertical focusing are essential.

The most recent beam physics and luminosity considerations for CLIC are presented in~\cite{PhysRevAccelBeams.23.101001}.
In a machine without imperfections, a vertical emittance of 6\,nm is achieved at the interaction point. The impact of static and dynamic imperfections is studied in~\cite{PhysRevAccelBeams.23.101001}. The dominant imperfections are the static misalignment of beamline elements, and ground motion. Beam-based alignment is used to minimise the impact of static imperfections. The beam-based alignment procedure for CLIC outperforms its requirement, which leads to significantly less vertical emittance growth than budgeted. For the expected alignment imperfections and with a conservative ground motion model, 90\% of machines achieve a luminosity of $4.5\times 10^{34}\,\text{cm}^{-2}\text{s}^{-1}$ or greater. This is the value used in~\ref{t:CLIC_stages_main}. The average luminosity achieved is $5.6\times 10^{34}\,\text{cm}^{-2}\text{s}^{-1}$. 
On-going studies~\cite{Adli:ESU25RDR}, including of the damping ring emittance, main linac tuning bumps and beam-delivery system improvements, show that significant luminosity improvements are possible.
A start-to-end simulation of a perfect machine shows that a luminosity of $8.6\times10^{34}\,\text{cm}^{-2}\text{s}^{-1}$ would be achieved. 


Novel system designs and technologies in combination with beam-based tuning
minimise the effect of imperfections, and simulated performances provide full luminosity up to \SI{3}{TeV}.
While this is not strictly necessary for lower energy stages, where the performance specifications could be relaxed,
this choice avoids the need for these systems to be upgraded for the higher-energy stages.
Key technologies that have been developed include the pre-alignment and quadrupole stabilisation systems,
as well as high-precision beam instrumentation.
Pre-alignment at the
\SI{10}{\micro\meter}-level is required for the main linac and BDS components.
An active alignment system achieves this, using actuators and sensors to remotely align the
\SI{2}{\meter}-long modules that support the main linac components 
with respect to a stretched-wire reference network. The alignment system has been successfully tested.
The disks that make up the accelerating structures are fabricated in
industry to micron tolerance and are bonded together to a tolerance of ten microns.
To mitigate the effect of wakefields caused by misalignments, the accelerating structures are equipped
with monitors that measure this wakefield to
determine the offset of the structure from the beam. The remote alignment system uses this information to minimise the
wakefield effects on the beam.
Nm-level vibration stabilisation systems have been developed for both main linac and final focus quadrupoles. The latter is based on actuators and inertial sensors, and allows decoupling of the focusing magnets from the ground.
This is important to avoid the natural ground motion as well as vibrations induced by
technical equipment that could lead to beam jitter and reduced luminosity.
Ground motion has been measured at CERN, and CLIC is designed to withstand the 
noise that has been measured in the CMS detector cavern.
Other key technologies that have been developed include hybrid, high-gradient final quadrupoles,
and advanced wigglers for the damping rings. 


%
%

In addition to hardware tests, a number of beam experiments provide the evidence that CLIC can reach its performance goals and that parameters can be met. A detailed description of key demonstrations are found in~\cite{Adli:ESU25RDR}. Some examples include:
 
\begin{itemize}
\item \textbf{Drive-Beam Scheme:} CTF3 has validated acceleration, RF transfer efficiency, power extraction, and two-beam acceleration up to 145 MV/m, with the required stability achieved~\cite{Corsini:2289699}.  
\item \textbf{Emittance \& focusing:} Modern light sources, such as the Swiss and Australian Light Sources~\cite{c:ls1,c:ls2,c:ls3}, achieve CLIC-level vertical emittances. Strong focusing has been tested at FFTB~\cite{Balakin1995}, ATF2~\cite{Kuroda2016,Okugi2016}, and the KEK B-factory~\cite{Thrane2017}.  
\item \textbf{Alignment \& Stability:} Beam-based alignment has been tested in FACET~\cite{FACET} and FERMI~\cite{FERMI}. CLIC's precision pre-alignment~\cite{Latina2014,Latina2014a} and active stabilization reduce jitter to the sub-nanometer regime~\cite{PhysRevAccelBeams.19.011001}.
\item \textbf{Timing \& Reliability:} CTF3 has demonstrated phase monitoring with fast correction~\cite{Corsini:2289699}. FELs provide long-range timing references, and high-availability concepts are informed by light sources, FELs, B-factories, and LHC operations.  
\end{itemize}


\paragraph{Construction, cost estimate, and energy consumption}

\begin{figure}[h!]
\centering
\includegraphics[width=0.7\textwidth]{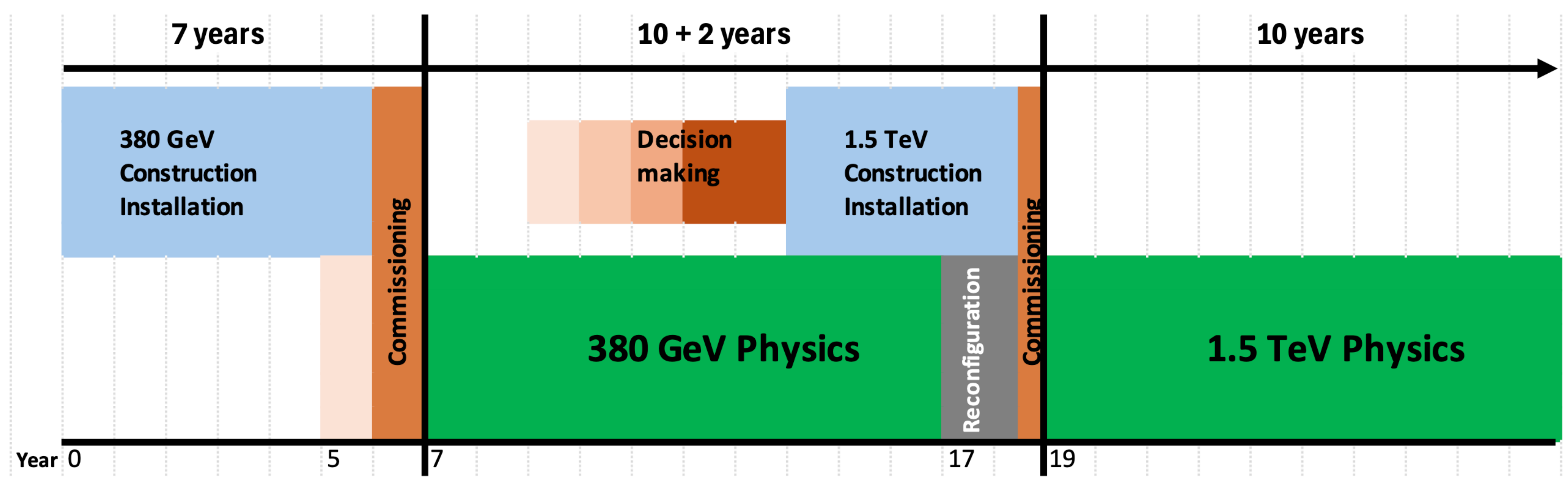}
\caption{\label{fig_IMP_9_main} Technology and construction-driven CLIC schedule, showing the construction and commissioning period and two stages for data taking. In the CLIC implementation planning the construction phase is increased to ten years to provide some contingency in the overall schedule presented. 
The time needed for reconfiguration (connection, hardware commissioning) between the stages is also indicated. The energy of the high-energy stage can be decided upon, and most of the tunnel can be dug and equipped, while running the initial stage.  \imcl}
\end{figure}

The technology and construction-driven timeline for the CLIC programme is given in~\ref{fig_IMP_9_main}~\cite{ESU18PiP}.
This schedule has seven years of initial construction and commissioning,
potentially starting $\sim$2034-2035.  Including a three-year margin, this leads to beam commissioning followed by first collisions a decade later.
The suggested 22 years of CLIC data-taking include an interval of two years between the stages.

\vspace*{.7cm}
\vspace*{.7cm}

The cost estimate of the 100 Hz initial stage with two IRs is
approximately \num{7.2}~billion~\si{CHF}.  The energy upgrade
to \SI{1.5}{\TeV} has an estimated cost of approximately \num{7.1}~billion~\si{CHF}, including the upgrade of the
drive-beam RF power.  

With the updated power consumption, see~\ref{fig_IMP_1}, the annual energy consumption for nominal running at the 100 Hz initial energy stage with two IRs
is estimated to be \SI{0.82}{TWh} and that of a high energy stage at 1.5\,TeV \SI{1.40}{TWh}.  For comparison, CERN's current energy consumption is
approximately \SI{1.2}{TWh} per year, of which the accelerator complex uses approximately 90\%.

\begin{figure}[htb!]
\centering
\includegraphics[width=1.0\textwidth, angle =0.0]
{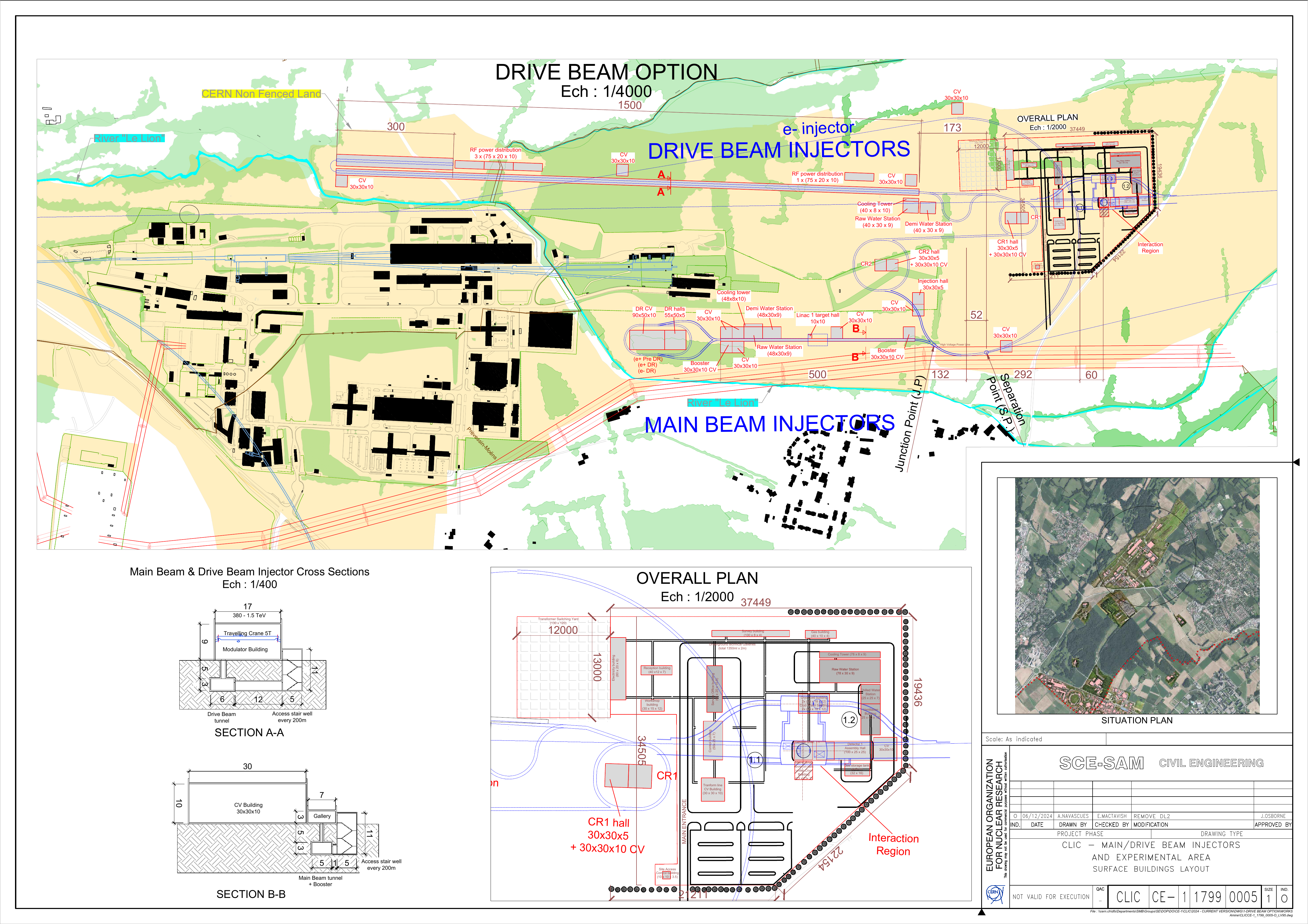}
\caption{\label{fig_CEIS_4} Schematic layout of the Injector and Experimental complexes, located the CERN Pr\'{e}vessin site. They have been optimised to be completely located on CERN land avoiding all existing CERN infrastructure and the key environmental concern of the river "Le Lion".}
\end{figure}

The layout of the injector and experimental complexes have been optimised to be completely located on CERN land, avoiding all existing CERN infrastructure and the key environmental concern of the river `Le Lion' (see ~\ref{fig_CEIS_4}). At 380\,GeV the two buildings housing the klystrons and modulators of the Drive-Beam Injector complex have a width of 17\,m and a combined length of 1800\,m. These buildings are situated on the west side of the Pr\'{e}vessin site. Surrounding these buildings is a series of CV, RF Power Distribution and Water Station buildings.

The Main-Beam Injectors are also located on the CERN site; they consist of several linacs and damping rings. The layout is similar to that presented in~\cite{ESU18PiP}. The Main-Beam Injector building has been reduced from a length of 880m to 500m and now includes the Booster in the same cut-and-cover tunnel beneath it. The Main-Beam Injectors have been relocated to the East of the Pr\'{e}vessin site, with the damping rings now being stacked. These optimisations significantly reduce both the number of required surface structures, and the transfer tunnel lengths within the complex.

Further details on the schedule, cost, power and energy and civil engineering are given in the Addendum and in~\cite{Adli:ESU25RDR}.

\eject

\section{CLIC Detector}
\label{sec:detector}
The design of the general-purpose CLIC detector is driven by the physics requirements and the stringent constraints imposed by the beam conditions \cite{cdrvol2}.
This section describes the requirements and the resulting optimized detector design, followed by a survey of recent detector-technology advancements towards reaching the ambitious performance goals.
\vspace{-0.5cm}
\paragraph{Detector requirements}
The demands for precision physics lead to challenging performance targets for the CLIC detector system.
A momentum resolution for high-momentum tracks of $\sigma_{\pT}/\pT^2 \leq \SI{2e-5}{\per\GeV}$ in the central detector is needed e.g. for the precision measurement of heavy states decaying into leptons. Optimal flavour tagging with clean b-, c-, and light-quark jet separation requires a resolution of \SI{5}{\um} in the transverse impact parameter, $d_0$, for single tracks in the central detector with $p_T$ above a few GeV. 
The jet energy resolution, $\sigma_E/E$, is required to be better than 5\% for light-quark
jets of 50~GeV, and better than 3.5\% for jet energies above 100~GeV. This is needed, for example, to separate W, Z
and H hadronic decays.  
A hit-time resolution of 1~ns in the calorimeters and 5~ns in the vertex- and tracking-detectors is needed, to efficiently suppress overlay of beam-induced background particles during the short bunch trains of 156--176~ns duration. Detector coverage for electrons and photons to very small polar angles (\SI{\sim10}{\mrad}) allows for a precise measurement of the luminosity and increases the physics reach for missing-energy signatures. The radiation-hardness requirements for the main detectors are very moderate, compared with high-luminosity hadron machines
\cite{ESU18RnD}. 
\vspace{-0.5cm}
\paragraph{Detector design}

\begin{wrapfigure}{r}{0.35\textwidth}
\vspace{-10pt}
    \includegraphics[width=\linewidth]{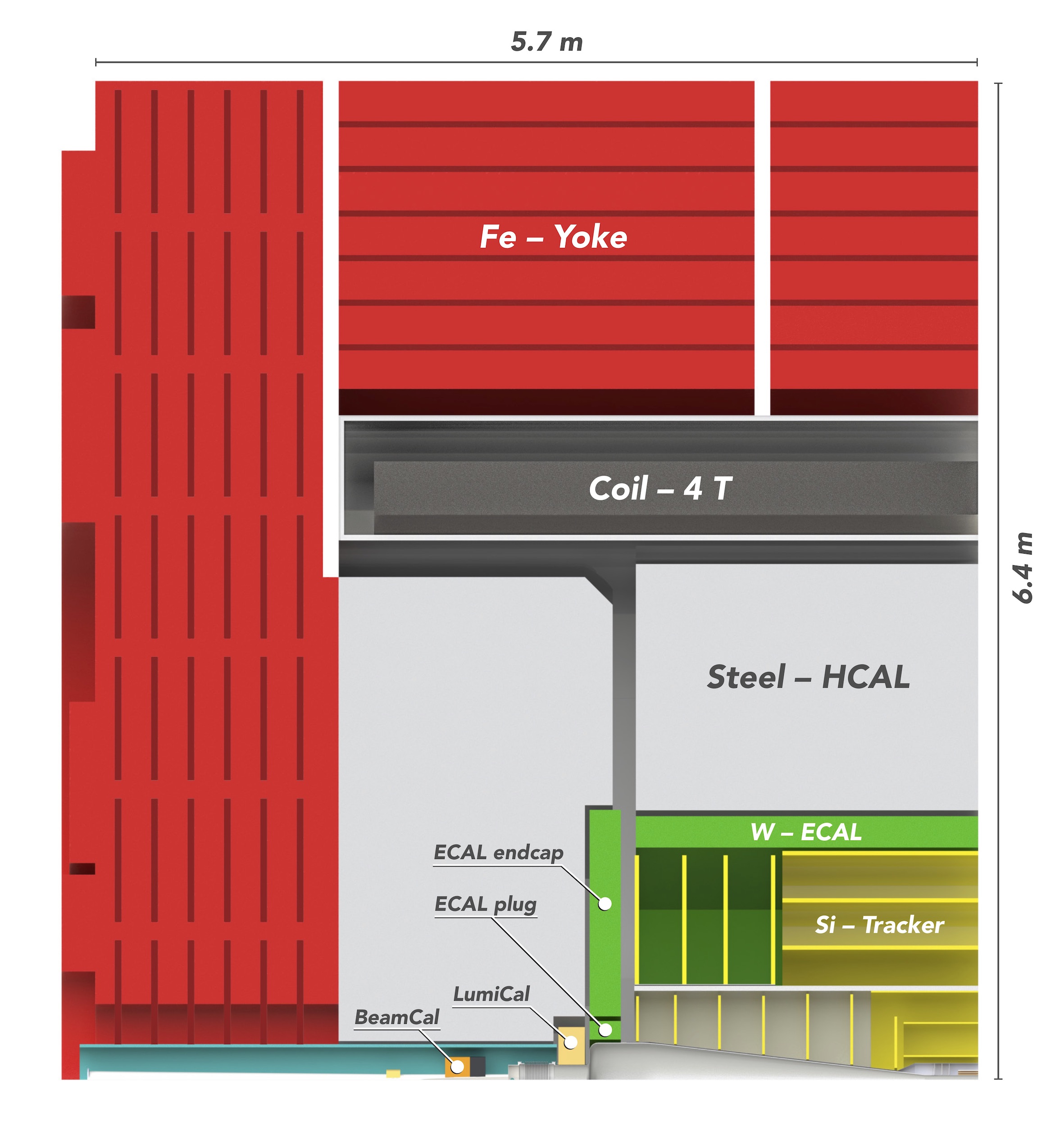}
  \caption{Longitudinal cross section of the top left quadrant of CLICdet.}\label{fig:clicdet}
  \vspace{-20pt}
\end{wrapfigure}

The optimized \textbf{CLICdet} detector concept \cite{CLICdet_note_2017,CLICdet_performance} is shown schematically in \cref{fig:clicdet}. 
Six silicon-pixel \textbf{vertex-detector} layers are foreseen, with a single-plane resolution of 3~\si{\micron} in both directions and a material budget of 0.2\% of a radiation length per layer, arranged in three double barrel layers and three spiral-shaped end-cap layers optimized for cooling with forced air flow. The \textbf{main tracker} contains six barrel and seven end-cap pixel-detector layers with a single-plane resolution in the $R-\phi$ direction of 7~\si{\micron}
for a total material budget of approximately 8\% of a radiation length in the barrel region.
The acceptance of the vertex- and tracking layers reaches down to track polar angles of approximately 7 degrees.
Accurate jet reconstruction through particle flow analysis (PFA) motivates the choice of
highly-granular \textbf{sampling calorimeters} with precise energy reconstruction and hit-time resolution. This implies densely integrated calorimetry with fully embedded electronics for a total of 110 million channels. The electromagnetic calorimeter (\textbf{ECAL}) of CLICdet is based on a 40-layer sandwich stack of silicon sensor pads and tungsten absorbers, while plastic scintillator tiles with SiPM readout and steel absorbers are proposed for the 60-layer hadron calorimeter (\textbf{HCAL}). 
Forward calorimeters located close to the beam pipe, called \textbf{LumiCal} and \textbf{BeamCal}, provide luminosity measurements and forward electron tagging. 
A superconducting \textbf{solenoid} provides a magnetic field of 4~T for the tracker and calorimeter volumes. It is surrounded by an iron yoke interleaved with detectors for muon identification.
\textbf{Triggerless readout} and \textbf{Power-pulsed} operation with the bunch-train repetition rate of 50--100~Hz are foreseen for all subsystems.

The performance of the CLICdet detector concept was successfully demonstrated in full-detector simulation studies covering centre-of-mass energies from 380~GeV up to 3~TeV, including event overlay from beam-induced backgrounds \cite{CLICdet_performance}. 
The recently updated CLIC accelerator parameters
for 380~GeV result in a 50\% increase in luminosity and a new operating scenario with a bunch-train repetition rate of 100~Hz instead of 50~Hz (see \cref{sec:accelerator}). The impact of this improved 380~GeV scenario on the detector has been assessed and found to be compatible with the CLICdet concept \cite{CLIC-lumi-bg:2025,CLIC-det-progress:2025}. The recent inclusion of a second interaction region in the CLIC accelerator design allows for hosting a second general-purpose experiment.
\vspace{-0.5cm}
\paragraph{Detector technology R\&D}
Ambitious R\&D programs for detector technologies meeting the requirements continue to be performed within the CLICdp collaboration and other collaborative frameworks \cite{AIDAinnova,EPreport2023,DRD3,OCTOPUS,DRD7,DRD6}, fully aligned with the 2021 ECFA Detector Roadmap \cite{ECFA-roadmap}. 

Initial results presented at the 2019 European Strategy Update Process demonstrated with proof-of-concept examples that the CLIC detector requirements were in reach with the advanced detector technologies pursued at that time \cite{ESU18RnD}.
Since then, the level of maturity of those technologies has increased substantially, thanks to detailed evaluations of constructed demonstrators.
Access to new and advanced industrial processes has enabled further progress towards reaching the performance requirements for CLIC \cite{CLIC-det-progress:2025}. This holds in particular in the hybrid and monolithic silicon-detector domain, which is crucial for the vertex and tracking detectors and for the ECAL. For example, a 65~nm CMOS monolithic process, already adapted and qualified for high-energy physics, will be explored for a staged sensor-development in view of the CLIC vertex-detector requirements \cite{OCTOPUS}. Progress in silicon-based timing detectors \cite{LGAD,FASTPIX_Braach_2023} opens up new applications of precise timing measurements in the CLIC detector, either as a dedicated timing layer
or integrated in the tracking layers.

Large-scale detector construction and upgrade projects serve as test-beds for the detector technologies foreseen for CLIC. These projects in progress include the High-Voltage CMOS sensors for the Mu3e experiment at PSI \cite{mupix_2024}, the upgrade of the ALICE inner tracking detector (ITS3) with 65~nm monolithic CMOS sensors \cite{ITS3-Aglietta_2024}, the new calorimeter endcaps of the CMS experiment at the HL-LHC \cite{CMShgcal-report2017}, and the calorimeter for the LUXE strong-field QED experiment at the European XFEL \cite{LUXE-TDR}. 

Flexible open-source hardware and software tools have been conceived in the context of the CLICdp project, to support detector R\&D and optimization studies: The \emph{Caribou DAQ} system for detector testing \cite{Caribou}, the \emph{Corryvreckan} versatile test-beam reconstruction and analysis framework \cite{Corryvreckan}, the \emph{Allpix-squared} silicon-detector simulation tool \cite{AllpixSquared,AllpixSquared_MAPS} and the \emph{Key4hep} turnkey software stack for detector optimization and performance studies for future collider experiments \cite{key4hep}. All tools are maintained and extended by large world-wide communities, enabling synergies across projects.

\section{Physics at CLIC}
\label{sec:physics}

\begin{figure}[bpht]
\centering
\begin{subfigure}{.43\textwidth}
\includegraphics[width=\linewidth]{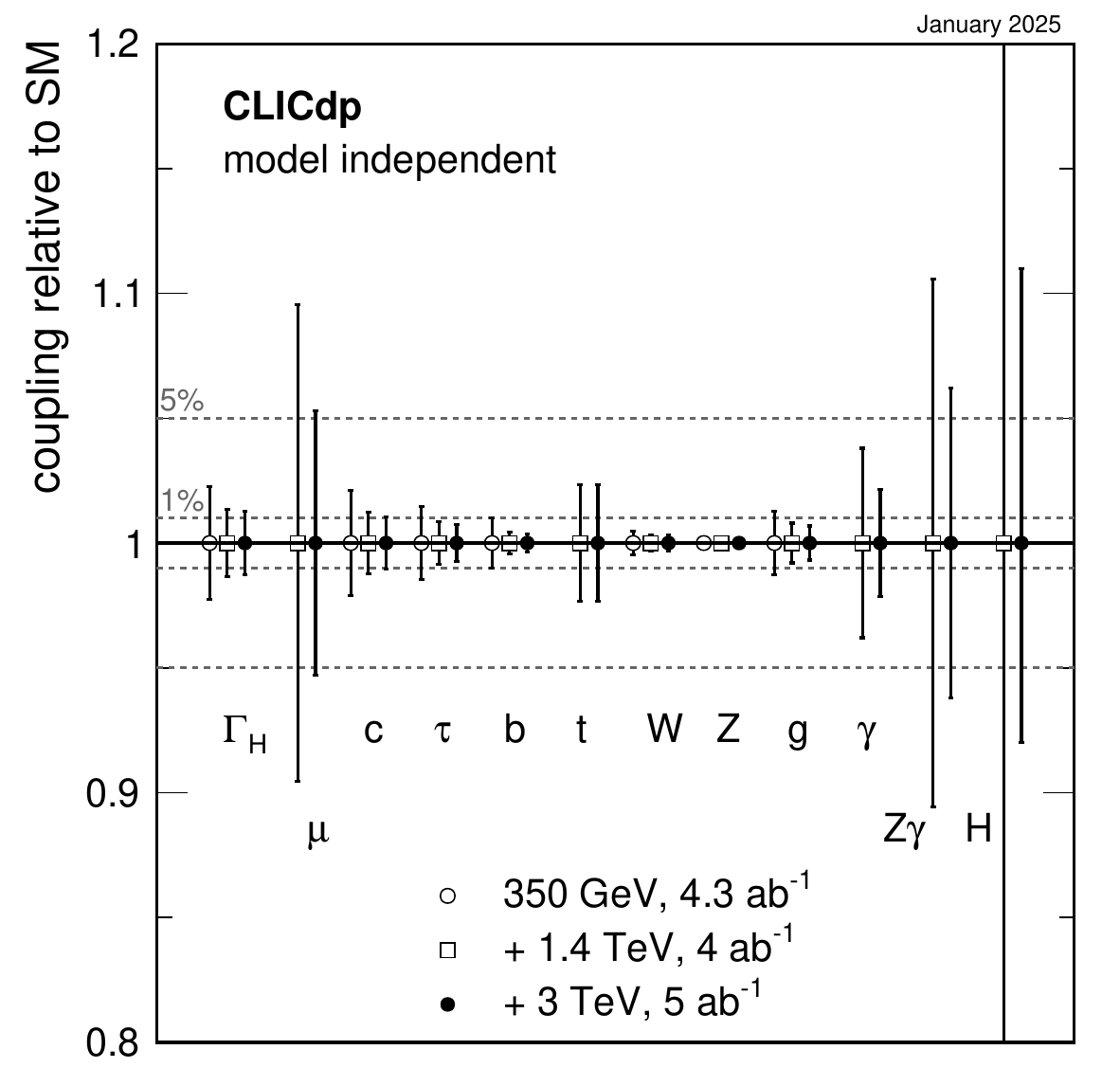}\label{fig:MIResultsPolarised8020_esu}
\end{subfigure}
\begin{subfigure}{.43\textwidth}
\includegraphics[width=\linewidth]{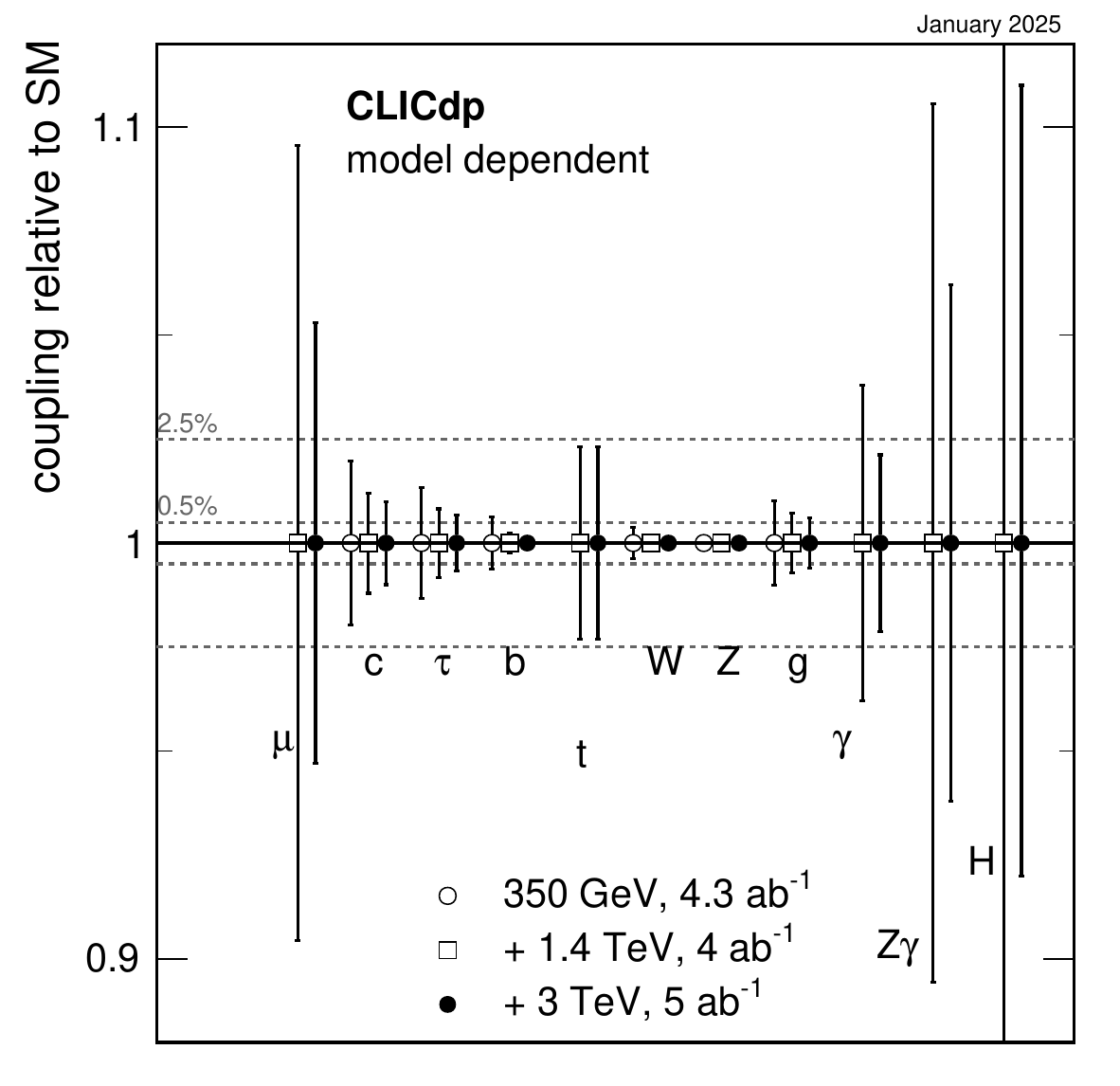}\label{fig:MDResultsPolarised8020_esu}
\end{subfigure}\vspace{-5pt}
\caption{CLIC results of (left) the model-independent fit and (right) the model-dependent fit to the Higgs couplings to SM particles. For the top-Higgs coupling, the $\SI{3}{\TeV}$ case has not yet been studied.}
\label{fig:HiggsResultsPolarised8020_esu}
\end{figure}

The CLIC physics programme will enable fundamentally new insights beyond the capabilities of the HL-LHC. The flexibility and very large accessible energy range provide a wide range of possibilities to discover new physics using very different approaches. The high centre-of-mass energy of CLIC extends the direct mass reach to scales far greater than that available at previous lepton colliders, surpassing even the HL-LHC for many signatures. The high luminosity and absence of QCD backgrounds give access to very rare processes at all energies. The clean experimental environment and absence of triggers in high-energy $\epem$ collisions and the good knowledge of the initial state allow precise measurements of many reactions to be performed, which probe the effects of new physics at mass scales far beyond the kinematic reach for direct production of new particles. The use of longitudinal electron beam polarisation enhances this reach further and may help to characterise newly discovered phenomena. Threshold scans provide very precise measurements of known particle masses. The CLIC experimental environment is also well-suited for looking for non-standard signatures such as anomalous tracks, peculiar secondary vertices, or unexpected energy depositions in the calorimeters.
\vspace{-0.5cm}
\paragraph{Higgs boson couplings}
The measurements of Higgs boson production cross sections times branching fractions in all accessible channels can be combined to extract the Higgs couplings and width. A model-independent global fit, described in \cite{ClicHiggsPaper}, makes use of the total cross section for the Higgsstrahlung process measured using the recoil mass method at the first energy stage to avoid any assumptions about additional BSM decays. For this reason, the initial CLIC stage is crucial for the Higgs physics programme. The results of the fit for the three CLIC energy stages are shown
in \ref{fig:HiggsResultsPolarised8020_esu}(left). While the results presented here are based on earlier studies that assumed an energy staging of $\roots = \SI{350}{\GeV}$, $\SI{1.4}{\TeV}$, and $\SI{3}{\TeV}$, the projections for these energy stages are scaled to the updated integrated luminosities~\cite{RobsonNoteHiggs25}. As explained in \autoref{sec:introduction}, this represents a factor 3 increase in luminosity performance at the first stage and foresees operation for 10 years with a 100 Hz bunch-train repetition rate at the first stage followed by 10 years operation at the second stage.

The expected precision on $g_\text{HZZ}$ is 0.3\% from the total ZH cross section at the initial stage, in a model-independent approach. Other couplings such as $g_\text{HWW}$ and $g_\text{Hbb}$ reach similar precision in the model-independent approach. The $g_\text{Hcc}$ coupling, which is very challenging at hadron colliders, can be probed with percent-level precision. The total Higgs width is extracted with 1.4\% accuracy using data from the first two stages.

Results from a global fit under the assumption of no non-SM Higgs boson decays~\cite{ClicHiggsPaper, RobsonNoteHiggs25}, which is model-dependent and equivalent  to the approach at hadron colliders, are illustrated in \ref{fig:HiggsResultsPolarised8020_esu}\,(right). In this case, several Higgs couplings are constrained to per mille-level precision at the high-energy stages.

\vspace{-0.5cm}
\paragraph{Higgs self-coupling}

The second CLIC energy stage allows a $5\,\upsigma$-observation
of the double Higgsstrahlung process e$^+$e$^-$ $\to$ ZHH and provides evidence for the WW fusion process e$^+$e$^-$ $\to$ HH$\nu_\text{e}\bar{\nu}_\text{e}$ assuming the SM value of $\lambda$. At $\roots=\SI{3}{\TeV}$, WW fusion is the leading double-Higgs boson production mechanism. The cross section is large enough to enable the measurement of differential cross sections to improve the knowledge of the Higgs self-coupling further. Using WW fusion and e$^+$e$^-$ $\to$ ZHH at the second stage and differential distributions for HH$\nu_\text{e}\bar{\nu}_\text{e}$ production at 3\,TeV leads to an expected precision on the Higgs self-coupling $\lambda$ of $[-8\%, +11\%]$~\cite{Roloff:2019crr}.
The inclusion of the ZHH cross section and the use of differential distributions avoids an ambiguity that occurs in the extraction of $\lambda$ from the HH$\nu_\text{e}\bar{\nu}_\text{e}$ cross section alone. Given the possible sizes of deviations in relevant extensions of the SM, the Higgs self-coupling measurement is an important motivation for CLIC operation in the multi-TeV region.
\vspace{-0.5cm}
\paragraph{Top-quark physics}

\begin{wrapfigure}{R}{0.45\textwidth}
\vspace{-10pt}
\includegraphics[width=\linewidth]{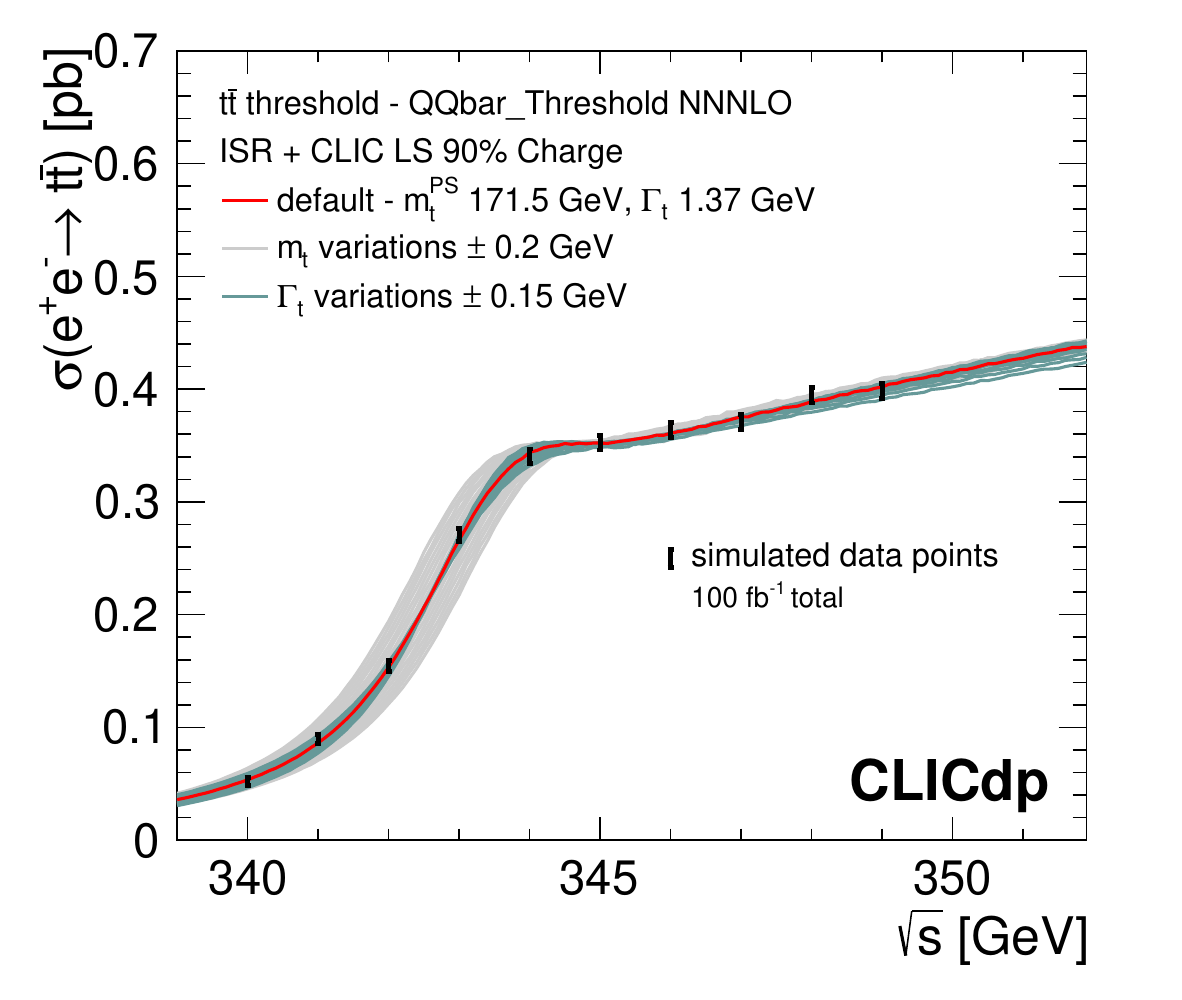}
\vspace*{-8mm}
\caption{Illustration of a top-quark threshold scan at CLIC, as explained in the text with a total integrated luminosity of \SI{100}{\per\fb}.}
\label{fig:TopThresholdScan_esu}
\vspace{-15pt}
\end{wrapfigure}

The top quark is the heaviest known fundamental particle and plays an important role in
many BSM theories; it therefore provides unique opportunities to test the SM and probe signatures of BSM effects. Already the first CLIC stage provides an important set of measurements using the e$^+$e$^-$ $\to$ t$\bar{\textnormal{t}}$ process.
A theoretically well-defined top-quark mass measurement can be performed in a threshold scan. Such a threshold scan is shown in \ref{fig:TopThresholdScan_esu}. The bands around the central cross section curve show the dependence of the cross section on the top-quark mass and width, illustrating the sensitivity of the threshold scan. The error bars on the simulated data points show the statistical uncertainties of the cross section measurement.
The pair production and decay of the top quark can be studied at 380\,GeV. The higher-energy stages provide complementary information on t$\bar{\textnormal{t}}$ production. Additionally, high-energy CLIC operation gives access to the t$\bar{\textnormal{t}}$H final state and to top-quark pair-production in vector boson fusion.

CLIC's prospects for top-quark physics have been studied with full detector simulation and are described in detail in \cite{CLICdp:2018esa}.
\vspace{-0.5cm}
\paragraph{New physics searches}

CLIC offers a rich potential for extensive exploration of the terascale in the form of direct and indirect searches of BSM effects. Direct searches are often possible up to the kinematic limit for particles with electroweak-sized coupling strength and detectable decay products. New physics effects, for example from scalars in an extended Higgs sector or from a composite Higgs sector, can be found directly or, beyond the kinematic reach, through effects of their mixing with known particles measured in Higgs boson production at CLIC. Long-lived charged particles such as the charged component of minimal dark matter multiplets can give rise to disappearing tracks, for which the clean environment and the detector layout are well suited. The measurement of double Higgs boson production will constrain models of electroweak baryogenesis. Other signatures include the measurement of soft decay products of new particles, e.g. from hidden sectors, which benefits from triggerless running and the clean environment. Additionally, high statistics of top quarks and Higgs bosons allow the search for rare decays indicating for example flavour violation effects. Direct and indirect searches for TeV-scale mediators of neutrino mass generation can provide further insights into the physics of flavour.

The CLIC potential to explore concrete new physics scenarios, which address several of the fundamental open questions of particle physics, is well documented in the literature. An overview is given in \cite{ESU18BSM}.
\section{Summary and Outlook}
\label{sec:summary}

CLIC is a mature proposal for the next generation of high-energy collider. Integrated beam simulations have concluded on a 50\% increase in luminosity with respect to the 2018 design. A significant power reduction has been achieved, enabling 100 Hz operation with a power consumption of \SI{166}{\mega\watt} at \SI{380}{\GeV}. This gives roughly three times higher luminosity-per-power with respect to 2018. An interaction region with two beam delivery systems hosting two detectors has been designed, where luminosity can be delivered to each detector. A higher-energy stage, still using the initial single drive-beam complex, can be optimized for any energy up to 2 TeV. Parameters are worked out in detail for a 1.5 TeV stage, with a site length of 29 km.  The precise energy of the high-energy stage can be decided upon, and most of the tunnel can be dug and equipped, while running the low-energy stage. The layout of the Injector and Experimental complexes, located at the CERN Pr\'{e}vessin site, has been optimised to be completely located on CERN territory.

The construction of the first CLIC energy stage could start as early as $\sim$2034-2035.  The preceding preparation phase is divided into two; in 2026-2028 the main technology R\&D will be completed, and the technical design for two IPs finalised.  After a process to validate the progress and promise of the project, site preparation, implementation studies, industrialisation of key components, and engineering design will start. See~\ref{tab:timeline_CLIC} for a detailed timeline.



CLIC offers the unique combination of high collision energies and the clean environment of e$^+$e$^-$ collisions. This enables the guaranteed physics programme of SM parameter measurements with unprecedented precision, ranging from the top-quark mass and other top-quark properties to the Higgs couplings, including the Higgs self-coupling. In addition, CLIC offers a rich potential for extensive exploration of the terascale in the form of direct and indirect searches of BSM effects. Direct searches are often possible up to the kinematic limit for particles with electroweak-sized coupling strength and detectable decay products. Beyond the kinematic reach, new physics effects might be found through effects of their mixing with known particles.

The physics goals and beam conditions at CLIC are addressed by an optimized CLICdet detector design.
Dedicated studies of innovative detector technologies for CLIC, as well as advancements in technologies available from industry and developed in synergy with other high-energy physics projects, have enabled significant progress towards reaching the stringent CLIC detector requirements. Large-scale detector systems currently in construction serve as test-beds for the CLIC detector concepts.

\newpage

\printbibliography[title=References]
\clearpage
\appendix
\section{Addendum}

\ref{addendum:stages} to \ref{addendum:status} contain answers to the "Questions for projects" defined by the Strategy Group.
\ref{addendum:community} gives information on the CLIC collaborations.


\subsection{Stages and Parameters}
\label{addendum:stages}

\subsubsection{The main stages of the project and the key scientific goals of each}

A baseline initial energy stage, starting at 380\,GeV, and a high-energy stage, currently designed at 1.5\,TeV.  The energy of the high-energy stage can be chosen after starting operation of the initial energy stage. Other energies are equally feasible.  250\,GeV or 550\,GeV can also be chosen for the first stage or in the case of 550\,GeV, as an upgrade from 380\,GeV. During the running of the initial stage, operation on the Z-pole is also possible. 

The initial energy stage at 380 GeV provides a compelling programme of precision Higgs and top-quark physics.  The dominant Higgs production mechanism at 380 GeV is the Higgsstrahlung process, while the vector-boson-fusion channel also contributes and provides complementary information; together, these enable a programme of precision Higgs coupling measurements.  The top quark has so far only been produced in hadron collisions; measurements in e$^+$e$^-$ collisions with different beam polarisations allow coupling contributions to be disentangled; and a theoretically well-defined top-quark mass measurement would be performed in a threshold scan around 350 GeV.  Higher-energy stages improve the Higgs precision programme via increased vector-boson-fusion production and provide unique sensitivity for a large number of new physics scenarios, as well as giving access to the Higgs self-coupling through double-Higgs production.


\subsubsection{Whether the ordering of stages is fixed or whether there is flexibility}

A high-energy stage comes after the initial 380\,GeV stage, and its construction can be done to a large degree in parallel with the operation of the initial stage. 

\subsubsection{For each stage, the main technical parameters}

\begin{table}[!htb]
\caption{Key parameters for 380\,GeV and 1.5\,TeV stages of CLIC.  Parameters for energy options at 250\,GeV and 550\,GeV are also given; for these options the power and luminosity are scalings, based on the 380\,GeV and 1.5\,TeV designs.
\newline
$^*$The luminosity for the 1.5 TeV machine has not been updated to reflect recent alignment studies~\cite{PhysRevAccelBeams.23.101001}. If the same method is applied, the luminosity at 1.5\,TeV is expected to reach 5.6 \SI{e34}{\per\centi\meter\squared\per\second}. }
\label{t:CLIC_stages}
\centering
\begin{tabular}{llllll}
\toprule 
Parameter  & Unit  & \textbf{380\,GeV}  &\textbf{1.5\,TeV}  & 250\,GeV  & 550\,GeV \tabularnewline
\midrule 
Centre-of-mass energy  & \si{\GeV}  & 380  & 1500  & 250  & 550 \tabularnewline
Repetition frequency  & \si{\Hz}  & 100  & 50 & 100  & 100 \tabularnewline
Nb. of bunches per train  &  & 352  & 312 & 352  & 352 \tabularnewline
Bunch separation  & \si{\ns}  & 0.5  & 0.5 & 0.5  & 0.5 \tabularnewline
Pulse length  & \si{\ns}  & 244  & 244 & 244  & 244 \tabularnewline
\midrule 
Accelerating gradient  & \si{\mega\volt/\meter}  & 72  & 72/100 & 72  & 72 \tabularnewline
\midrule 
Total luminosity  & \SI{e34}{\per\centi\meter\squared\per\second}  & 4.5  & 3.7$^{*}$  & $\sim$3.0  & $\sim$6.5 \tabularnewline
Lum. above \SI{99}{\percent} of $\sqrt{s}$  & \SI{e34}{\per\centi\meter\squared\per\second}  & 2.7  & 1.4  &  $\sim$2.1  &  $\sim$3.2 \tabularnewline
Total int. lum. per year  & fb$^{-1}$  & 540  & 444  & $\sim$350  & $\sim$780 \tabularnewline
Power consumption  & MW  & 166  & 287  & $\sim$130  & $\sim$210 \tabularnewline
\midrule 
Main linac tunnel length  & \si{\km}  & 11.4  & 29.0 & 11.4  & $\sim$15 \tabularnewline
Nb. of particles per bunch  & \num{e9}  & 5.2  & 3.7 & 5.2  & 5.2 \tabularnewline
Bunch length  & \si{\um}  & 70  & 44 & 70  & 70 \tabularnewline
IP beam size  & \si{\nm}  & 149/2.0  & 60/1.5 & $\sim$184/2.5  & $\sim$124/1.7 \tabularnewline
\bottomrule
\end{tabular}
\end{table}

\subsubsection{The number of independent experimental activities and the number of scientists expected to be engaged in each}

Two multi-purpose detectors can be installed each with a number of scientists similar to the LHC experiments. CLIC, like other Higgs-factories, can also support additional smaller experimental collaborations making use of the injectors or extracted beams (e.g. as discussed in the Physics Beyond Collider submission to the ESPP).

\subsection{Timeline}

\subsubsection{The technically-limited timeline for construction of each stage}

\begin{figure}[h!]
\centering
\includegraphics[width=0.50\textwidth]{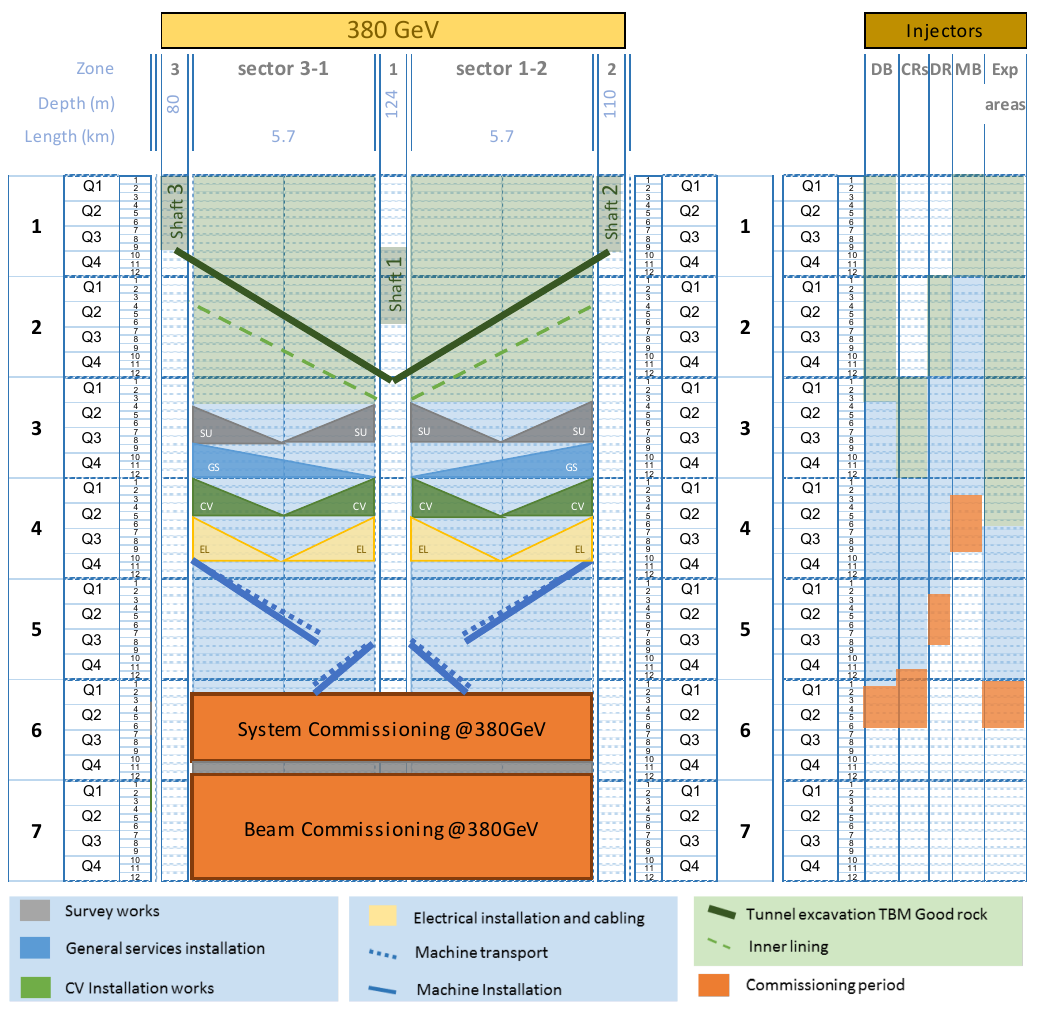}
\caption{\label{fig_IMP_6} Technically limited construction and commissioning schedule for the 380\,GeV drive-beam based CLIC facility. 
The vertical axis represents time in years. 
}
\end{figure}

The construction, installation, and commissioning schedule for the \SI{380}{\GeV} CLIC complex is shown in~\ref{fig_IMP_6} and lasts for 7 years. It comprises:

\begin{itemize}
\item  Slightly more than five years for the excavation and tunnel lining, the installation of the tunnel infrastructures, and the accelerator equipment transport and installation.
\item  Eight months for the system commissioning, followed by two months for final alignment.
\item  One year for the accelerator commissioning with beam.
\end{itemize}



\begin{figure}[h!]
\centering
\includegraphics[width=0.7\textwidth]{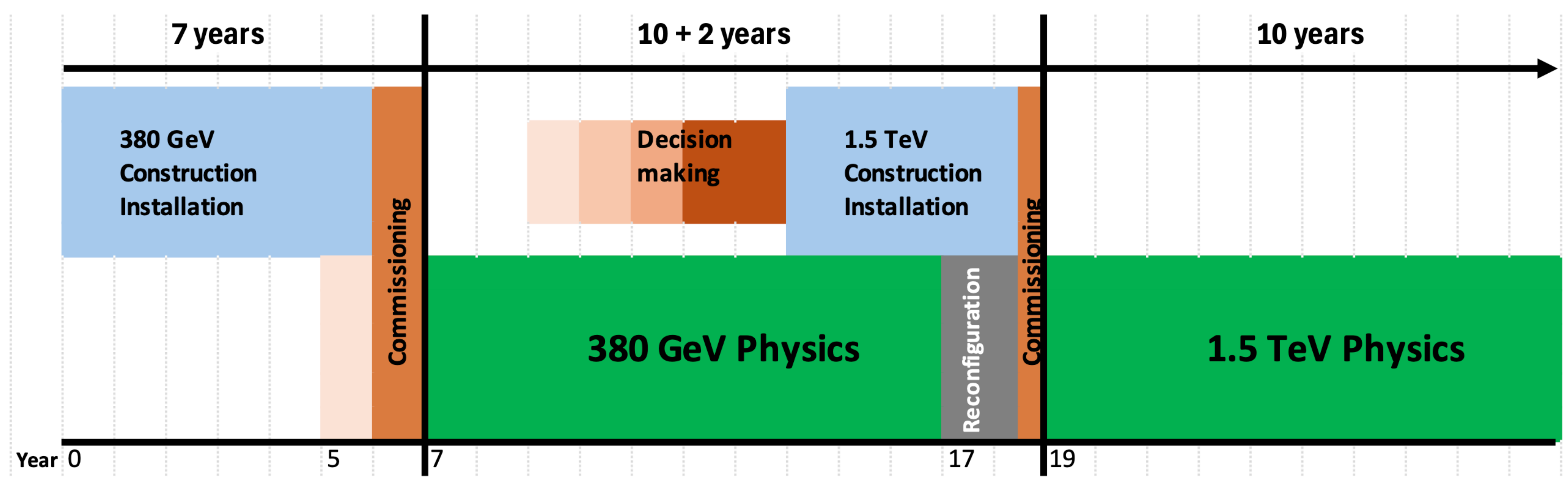}
\caption{\label{fig_IMP_9} Technology and construction-driven CLIC schedule, showing the construction and commissioning period and two stages for data taking. In the CLIC implementation planning the construction phase is increased to ten years to provide some contingency in the overall schedule presented in ~\ref{tab:timeline_CLIC}. 
The time needed for reconfiguration (connection, hardware commissioning) between the stages is also indicated and will be further refined based on the experience gained during the initial stage installation and operation phases. The energy of the high-energy stage can be decided upon, and most of the tunnel can be dug and equipped, while running the first energy stage. }
\end{figure}

In a schedule driven by technology and construction, the CLIC project would cover 29 years as shown in ~\ref{fig_IMP_9}~\cite{ESU18PiP}, counted from the start of construction. This schedule has a technically limited seven years of construction and commissioning phase, which we increase to ten years to provide some contingency in the overall schedule presented in~\ref{tab:timeline_CLIC}. The suggested 22 years of CLIC data-taking include an interval of two years between the stages. 



\vspace*{.7cm}
\vspace*{.7cm}

\subsubsection{The anticipated operational (running) time at each stage, and the expected operational duty cycle}

\begin{figure}[!htb]
\begin{flushleft}
\hspace{1.5cm}
\begin{adjustbox}{width=0.6\linewidth}
\begin{tikzpicture}[font=\sffamily,lines/.style={draw=none},]
\sansmath
\pie [rotate = 90,
scale font=false,
radius = 4,
text = legend,
sum=auto,
every only number node/.style={text=white},
style={lines},
    color={
    red!60,
    orange!60,
    yellow!60,
    red!60!blue!60,
    red!60!cyan!60,
    cyan!60!yellow!60
    },
] {
120 / Annual shutdown,
30 / Commissioning,
10 / Technical stops,
20 / Machine development,
46 / Fault induced stops,
139 / Data taking
}
\end{tikzpicture}
\end{adjustbox}
\end{flushleft}
\caption{\label{fig_IMP_12} Operation schedule in a "normal" year (days/year).}
\end{figure}
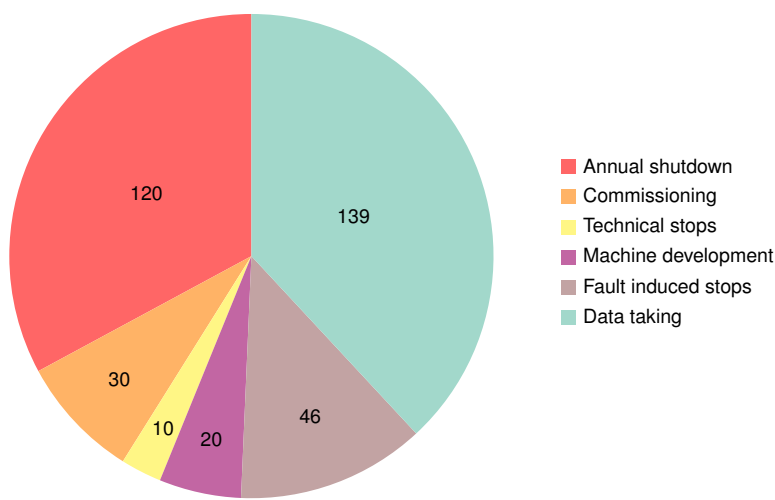

The operational scenario is detailed in~\cite{Bordry:2018gri} and is depicted in~\ref{fig_IMP_12}.

The baseline plan for operating CLIC includes a yearly shutdown of 120 days, 30 days of commissioning,
20 days for machine development, and 10 days for planned technical stops.
This leaves 185 days of operation for collider physics data-taking. Assuming an availability during normal running of 75\% (i.e. 139 days running and 46 days fault induced stops), this results in an integrated
luminosity per year equivalent to operating at full luminosity for \SI{1.2e7}{\second}~\cite{Bordry:2018gri}.

\subsection{Resource Requirements}

\subsubsection{The capital cost of each stage in 2024 CHF}
\label{addendum:costings}

The initial energy stage of CLIC, as well as delta costs of upgrading to higher energies, have been fully re-costed in 2024 CHF (with 1 EURO = 0.94 CHF). For the cost estimate, a bottom-up approach is used, following the work breakdown structure of the project, starting from unit costs and quantities for components, and then moving up to technical systems, subdomains and domains.
For some parts (e.g. standard systems), cost scaling from similar items is used.

The cost estimates cover the project construction phase, from approval to start of commissioning with beam. 
They include all the domains of the CLIC complex from injectors to beam dumps, together with the corresponding civil engineering and infrastructures. 
Items such as R\&D, prototyping and pre-industrialisation costs, acquisition of land and underground rights-of-way, computing, and general laboratory infrastructures and services (e.g.\ offices, administration, purchasing and human resources management) are excluded.

The breakdown of the resulting cost estimate up to the sub-domain level is presented in~\ref{Tab:Cost} for the 100 Hz, double BDS-version of the 380\,GeV stage of the accelerator complex. 
The injectors for the main-beam and drive-beam production are among the most expensive parts of the project, together with the main linac, and the civil engineering and services.

\clearpage
\begin{table}[ht]
\caption{Cost breakdown for the 380\,GeV stage of the CLIC accelerator.}
\label{Tab:Cost}
\centering
\begin{tabular}{l l S[table-format=4.0]}
\toprule
Domain & Sub-Domain & Cost [\si{MCHF}] \\
\midrule
 \multirow{3}{*}{Main-Beam Production} & Injectors & 168 \\
 & Damping Rings & 386 \\
 & Beam Transport & 492 \\ 
 \midrule
\multirow{2}{*}{Drive-Beam Production} & Injectors & 560 \\
 & Frequency Multiplication and Beam Transport & 500 \\ \midrule
\multirow{2}{*}{Main Linac Modules}  & Main Linac Modules & 2133 \\
 & Post decelerators  & 46 \\ \midrule
\multirow{3}{*}{\makecell[l]{Beam Delivery and \\ Post Collision Lines}}   & Beam Delivery Systems & 172 \\
 & Final focus, Exp. Area & 20 \\
 & Post-collision lines/dumps & 100 \\ \midrule
\multirow{2}{*}{Civil Engineering} & Surface buildings & 367 \\
& Underground structures & 936 \\ \midrule
\multirow{4}{*}{Infrastructure and Services}  & Electrical distribution  & 265 \\
 & Survey and Alignment & 204  \\
 & Cooling and ventilation  & 562 \\
 & Transport / installation & 62\\ \midrule
\multirow{4}{*}{\makecell[l]{Machine Control, Protection \\ and Safety systems}} & Safety systems  & 76 \\
  & Machine Control Infrastructure & 153 \\
 & Machine Protection & 15 \\
 & Access Safety \& Control System & 24  \\ \midrule
\bfseries Total (rounded) & & \bfseries 7240  \\
\bottomrule
\end{tabular}
\end{table}


Combining the estimated technical uncertainties yields a total (1$\sigma$) error of 1561\,MCHF for the 380\,GeV facility.
In addition, the commercial uncertainties, defined in Section 6 of~\cite{Adli:ESU25RDR}, need to be included. They amount to 901\,MCHF.
The total uncertainty is obtained by adding technical and commercial uncertainties in quadrature. Finally, for the estimated error band around the cost estimate, the resulting total uncertainty is used on the positive side, while only the technical uncertainty is used on the negative side~\cite{cdrvol3}. The cost estimate for the first stage of CLIC including a 1$\sigma$ overall uncertainty is therefore:
\begin{center}
\begin{tabular}{lcc}
CLIC 380 GeV, 100 Hz, two interaction regions &:& $7240^{+1807}_{-1561}$\,MCHF\quad.\\ \\
\end{tabular}
\end{center}

\begin{figure}[h]
\begin{center}
\includegraphics[width=13 cm]{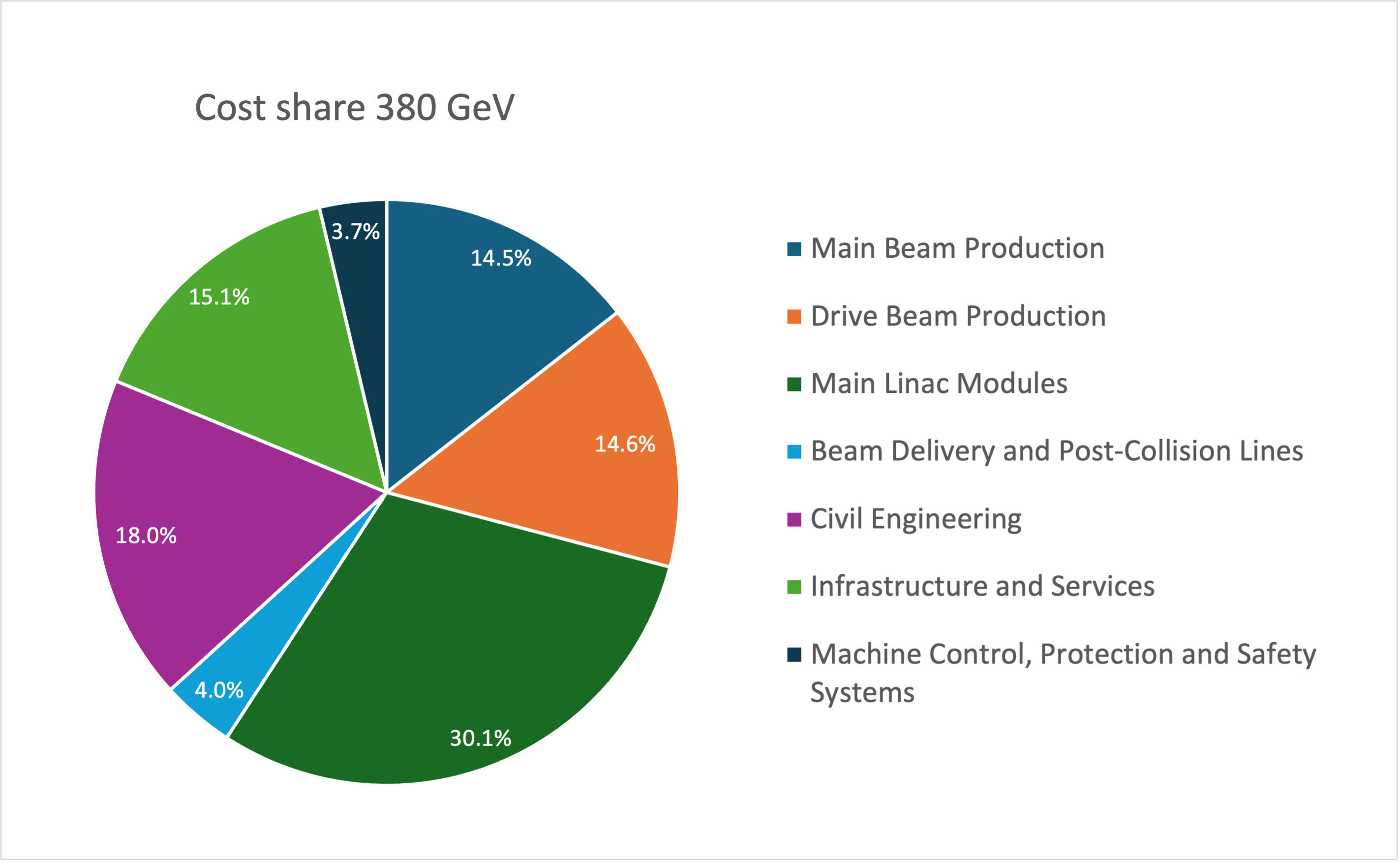}
\caption{Cost breakdown for the 100 Hz 380\,GeV stage of the CLIC accelerator, with a dual beam-delivery system and two interaction regions. The total cost is 7240\,MCHF.}
\label{fig:CLIC_CostShare}
\end{center}
\end{figure}

~\ref{fig:CLIC_CostShare} illustrates the sharing of cost between different parts of the accelerator complex. The injectors for the main-beam and drive-beam production are among the most expensive parts of the project, together with the main linac, and the civil engineering and services. A very preliminary estimate, based on scaling from other CERN projects and studies, of costs related to treatment of excavated materials would add 1.5-2\%.

The cost composition and values of the 1.5~TeV complex have also been estimated. The energy upgrade to 1.5\,TeV has a cost estimate of $\sim \text{7.1}\,\text{billion CHF}$, including the upgrade of the drive-beam RF power needed for the 1.5\,TeV stage.
The 250\,GeV configuration will cost around 10\% less than the 380\,GeV complex, and a 550\,GeV configuration around 30\% more.

The methodology used for estimating the cost of the CLIC detector~\cite{CLICdet_note_2017}
is similar to that used for the accelerator complex, and is based on the detector work breakdown structure. 
A breakdown of the value estimate for the CLIC detector is given in~\ref{Tab:Det_Cost}, and comes to a total of approximately \num{400}~million~\si{CHF}.
The main cost driver is the cost of the silicon sensors
for the 40-layer Electromagnetic Calorimeter (ECAL). For example, a 25\% reduction in the cost of silicon per unit of surface would reduce the overall detector cost by more than 10\%. Alternative designs for ECAL are feasible, but will reduce the detector performance (e.g. worse energy resolution for photons~\cite{CLICdet_note_2017}).

\begin{table}[ht]
\caption{Cost estimate of the CLIC detector~\cite{clicdet_cost}.}
\label{Tab:Det_Cost}
\centering
\pgfplotstableread[header=true]{
name z
Vertex 13
{Silicon Tracker} 43
{Electromagnetic Calorimeter} 180
{Hadronic Calorimeter} 39
{Muon System} 16
{Coil and Yoke} 95
{Other} 11
{ } {nan}
{Total } 397
}\data

\pgfplotstableset{create on use/error/.style={
    create col/expr={\thisrow{z}
    }
  }
}

\pgfplotsset{select coords between index/.style 2 args={
    x filter/.code={
        \ifnum\coordindex<#1\def\pgfmathresult{}\fi
        \ifnum\coordindex>#2\def\pgfmathresult{}\fi
    }
}}

\newcommand{\errplot}{%
\begin{tikzpicture}[trim axis left,trim axis right]
\begin{axis}[y=-\baselineskip,
xbar,
xmax=50,
width             = 7.5cm,
axis y line=none,
ytick             = \empty,
xtick={0,10,...,50},
xticklabels = {0,10\%,20\%,30\%,40\%,50\%},
axis x line*      = bottom,
]
\addplot+[xbar,fill=orange,draw=black,select coords between index={0}{7}] table [x expr ={\thisrowno{1}/397*100},y expr=\coordindex]{\data};
\end{axis}
\end{tikzpicture}
}

\pgfplotstablegetrowsof{\data}
\let\numberofrows=\pgfplotsretval

\pgfplotstabletypeset[columns={name,error,z},
  col sep = comma,
  every head row/.style = {before row=\toprule, after row=\midrule},
  every last row/.style = {after row=[0ex]\bottomrule, before row=[1ex]\midrule},
  columns/name/.style = {string type, column name=System},
      every row 7 column 2/.style={
        postproc cell content/.style={
          @cell content=\textcolor{white}{##1}
        }
      },
  columns/error/.style = {
    column name = {Cost fraction},
    assign cell content/.code = {
    \ifnum\pgfplotstablerow=0
    \pgfkeyssetvalue{/pgfplots/table/@cell content}
    {\multirow{\numberofrows}{6cm}{\errplot}}%
    \else
    \pgfkeyssetvalue{/pgfplots/table/@cell content}{}%
    \fi
    }
  },
  columns/z/.style    = {column name = {Cost [\si{MCHF}]},   column type={S[table-format=2.1]}, string type},
]{\data}

\end{table}

\subsubsection{The annual operation cost of the accelerator}

A preliminary estimate of the CLIC accelerator operation cost, including replacement/operation of
accelerator hardware parts, RF systems, and cooling, ventilation, electronics, and electrical infrastructures, 
amounts to 137\,MCHF per year. 
This is done by classifying construction items into:
\begin{itemize}
    \item Fixed accelerator installations, taken to be the total costs for modules, magnets, etc
    \item "Consumables", taken to be the costs for RF power equipment, magnet power supplies, etc
    \item Technical infrastructure, taken to be the costs for infrastructure such as cooling and ventilation, electric power infrastructure, installation equipment, etc

\end{itemize}
and assuming a replacement and maintenance cost per year of 1\,\% / 3\,\% / 5\,\%, respectively, of the corresponding capital expense.

An important ingredient of the operation cost is the CLIC power consumption and the corresponding energy cost. With an energy consumption at 0.82\,TWh for 380\,GeV and a price of 80\,CHF / MWh the energy costs will be 66\,MCHF annually, increasing to 112\,MCHF at 1.5 TeV for 1.40\,TWh.  

\subsubsection{The human resources (in FTE) needed to deliver or operate each stage over its lifetime, expressed as an annual profile}
The Snowmass 2021 Collider Implementation Task Force~\cite{Roser_2023} proposed a general formula for estimating the labour needed for construction: 
\begin{equation}
Explicit \, Labour  = 15.7 \cdot (Value)^{0.75},
\label{ModelLabor}    
\end{equation}
with Explicit Labour in FTE-years and Value in MCHF of 2010, excluding civil engineering. 
Applying the formula above yields approximately 10500 FTEy of explicit labour. An important part of these could be covered outside CERN, for in-kind deliverables. 

Concerning personnel needed for the operation of CLIC, one can assume efforts that are similar to large accelerator facilities operating today. Much experience was gained with operating Free Electron Laser linacs and light-sources with similar technologies.
The maintenance programme for equipment in the klystron galleries is demanding, but is not expected to impact strongly on the overall personnel required for operation. The RF systems, injectors and drivebeam, are above ground concentrated in a small area at the CERN Pr\'{e}vessin site. 
The ILC project has made a detailed estimate of the personnel needed to operate ILC, yielding 640\,FTE for 250 GeV and 850\,FTE for 500\,GeV. 
Such numbers are also in line with LHC experience, assuming personnel costs to be similar to hardware maintenance costs, and estimating an average cost of 210k per FTEy.
In the framework of CERN, these numbers would distribute across scientific/engineering/technical staff, technical service contracts, fellows and administrative staff. 
The level of CLIC annual operational personnel support required is thus expected to be around 650\,FTEy for CLIC 380 GeV.

\subsubsection{Commentary on the basis-of-estimate of the resource requirements}
This is covered above.

\subsection{Environmental Impact}

\subsubsection{The peak (MW) and integrated (TWh) energy consumption during operation of each stage}

The nominal power consumption at the 380\,GeV stage has been estimated based on the 
detailed CLIC work breakdown structure. This yields, for the 100\,Hz baseline option at \SI{380}{\GeV} with two interaction regions, a total of 166\,MW for all accelerator systems and services, taking into account network losses for transformation and distribution on site.  For the 50\,Hz reduced power machine, assuming one interaction region, the total is 105\,MW. 
The equivalent analysis at \SI{1.5}{\TeV} yields 287\,MW running at 50\,Hz. Most of the power is used in the drive-beam and main-beam injector complexes, comparatively little in the main linacs. Among the technical systems, the RF represents the major consumer.

\begin{table}[ht]
\caption{Estimated power consumption of CLIC at the two centre-of-mass energy stages and for different operation modes. 
}
\label{Tab:Power}
\centering
\begin{tabular}{S[table-format=4.0]S[table-format=3.0]S[table-format=2.0]S[table-format=2.0]}
\toprule
{Collision energy [GeV]}  & {Running [MW]} &  {Short [MW]} & {Off [MW]} \\
\midrule
380\,\, 50\,Hz     &  105  & 17  & 10 \\
380\,\, 100\,Hz    &  166  & 17  & 10 \\
1500\,\, 50\,Hz    &  287  & 33  & 13 \\
\bottomrule
\end{tabular}
\end{table}

\ref{Tab:Power} shows the nominal power consumption in three different operation modes of CLIC, including the "running" mode at the different energy stages, 
as well as the residual values for two operational modes corresponding to short ("short") and long ("off") beam interruptions. 

In terms of energy consumption the accelerator is assumed~\cite{Bordry:2018gri} to be "off" for 120 days and "running" for 139 days. The power consumption during the remaining time, covering commissioning, technical stops, machine development and fault-induced stops,
is taken into account by estimating a 50/50 split between "running" and "short" during these 106 days. 
The resulting electrical energy is shown in~\ref{Tab:Energy}. 

\begin{table}[ht]
\caption{Estimated annual power consumption of CLIC at the two centre-of-mass energy stages and for different operation modes. 
}
\label{Tab:Energy}
\centering
\begin{tabular}{S[table-format=4.0]S[table-format=3.0]}
\toprule
{Collision energy [\si{\GeV}]}  & {Annual Energy Consumption [TWh]} \\
\midrule
380\,\, 50\,Hz  &  0.54  \\
380\,\, 100\,Hz &  0.82  \\
1500\,\, 50\,Hz &  1.40  \\
\bottomrule
\end{tabular}
\end{table}



\subsubsection{The integrated carbon-equivalent energy cost of construction}

The ILC and CLIC projects have commissioned two lifecycle assessment (LCA) studies with ARUP, a consultancy company \cite{clic_ilc_lca_arup,clic_ilc_lca-accel_arup}. 
These lifecycle assessments form the basis for the Green House Gas (GHG) emission numbers summarised below, and also in~\ref{tab:ghg-project-CLIC}:

\begin{itemize}
   \item "Cradle to gate analysis" (LCA A1 to A5 modules) focussing on GHG emissions and therefore expressed in g CO$_2$ eq. (or multiples of) for:
    \begin{itemize}
        \item civil engineering;
        \item hardware components and infrastructure for the accelerator complex;
        \item detectors.
    \end{itemize}
    \item The Global Warming Potential (GWP) emission due to the generation of the electricity required for the operation of the accelerator complex (main collider and injectors).
\end{itemize}

Although the above list is not exhaustive, it provides an indicative assessment of the relative impact of construction and operation of the accelerator. In~\ref{tab:ghg-project-CLIC} several parameters are still being estimated but some conclusions can be made:

\begin{itemize}
\item For CLIC the embodied carbon in materials used for civil engineering and accelerator components is significantly larger (by a factor 3-4) than the emissions due to 10 years of operation. This factor will become smaller when the reductions possible for 2040 for the embodied carbon have been analysed and validated in more detail. 
\item The surface installations, drive-beam \& main-beam injectors and buildings for these, plus buildings at the integration points of the two experiments and near shafts, contribute a factor two-three more than the initial 11.5\,km main tunnel. This also means that layout changes and optimisation of cut-and-cover installations and related building can provide important additional improvements for CLIC.
\item While reduction of the impact of the civil engineering -- close to a factor two -- seems within reach on the timescale of 2040, the reductions due to decarbonising of the society/industry/materials as a whole on this timescale are difficult to estimate for the accelerator, detector and infrastructure components. 
\item First estimates of the impact of two detectors, even with supports and all their services not included, indicate that they contribute at a level approaching the civil engineering impact of 10\,km of tunnel. 
\end{itemize}

\begin{table}[!htbp]
 \small
 \centering
 \begin{tabular}{|l|c|c|c|}
 \hline 
   \textbf{ CLIC } & Stage 1 & Stage 2 & Reduce by 2040-50   \\
 \hline
    CoM energy [\unit{\GeV}] & 380 & 1500 & \\ \hline
    Luminosity/IP $[10^{34} \unit{\cm^{-2}}\unit{\s^{-1}}~]$ & 2.3 & 3.7 &  \\ \hline
    Number of IPs & 2 & 1  & \\ \hline
    Operation time for physics/yr  $[10^7 \unit{s}/ \unit{yr}]$ & \multicolumn{2}{c|}{1.2} &  \\ \hline
    Integrated luminosity/ \unit{yr}  [1/\unit{fb}/ \unit{yr}] & 540  & 444 &  \\ \hline
    Host countries & 
           \multicolumn{2}{c|}{France and Switzerland } & \\
 \hline 
 \multicolumn{4}{|l|}{\textbf{GHG emissions from construction, stage A1-A5 (CE), A1-A3 (HW)}} \\
 \hline 
    Subsurface structures [kt CO2 eq.] & 286  & 199 & 40-50\% \\ \hline 
    Surface sites constructions [kt CO2 eq.] & 118 & 0 & 40-50\% \\ \hline 
    Accelerator (coll.) [kt CO2 eq.] & 60 & $\sim$125 & 25\%\\
 \hline 
    Accelerator (drivebeam and inj.) [kt CO2 eq.] & 80 & $\sim$10 & 25\%\\
    \hline
Services [kt CO2 eq.] & 19 & $\sim$20 & 25\%\\
 \hline 
 
    Two detectors [kt CO2 eq.] & 94 & upgrades & 25\% \\
 \hline 
    \textbf{Total [kt CO2 eq.]} & \multicolumn{2}{c|}{ } & \\ \hline 
    Collider tunnel length  [\unit{km}] & 11.5 & 29.1 &  \\ \hline
    Collider tunnel diameter [\unit{m}] & \multicolumn{2}{c|}{5.6} & \\ \hline
    Collider tunnel GWP/m [t CO2 eq/\unit{m}] & \multicolumn{2}{c|}{8.1} & \\ \hline
    Concrete GWP  [kg CO2 eq./kg] & \multicolumn{2}{c|}{0.16} & \\
 \hline 
    Accelerator  GWP/m [t CO2 eq./m] & \multicolumn{2}{c|}{Not estimated} & \\
 \hline 
 \multicolumn{4}{|l|}{\textbf{GHG emissions from operation}} \\
 \hline 
    Maximum power in operation [\unit{MW}]   & 166 & 287 &  \\ \hline
    Average power in operation [\unit{MW}]   &  &  &  \\ \hline
    Electricity consumption / yr  [\unit{TWh}/ \unit{yr}] & 0.82 & 1.40 & \\ \hline
    Years of operation & 10 & 10  & \\ \hline
    Carbon intensity of electricity [g CO2 eq./ \unit{kWh}] & \multicolumn{2}{c|}{16} & Estimate for 2050\\ \hline
    Average Scope 2 emissions / yr [kt CO2 eq.] & 13 & 22 & Estimate for 2050 \\
 \hline 

 \end{tabular}

 \caption{Data on GWP for the CLIC project. These numbers are based on the reports mentioned above and will be further developed. Stage 2 is the delta for the upgrade based on the increase needed of the relevant components. All numbers are today's values, except the carbon intensity of electricity for operation. The last column indicate reduction potentials. }

 \label{tab:ghg-project-CLIC}

\end{table}

\subsubsection{Any other significant expected environmental impacts}

Roads and land acquisition will be addressed as part of future site and placement studies.
CLIC will be powered from the existing power substation infrastructure at the Pr\'{e}vessin site. 

\subsection{Technology and Delivery}

The implementation schedule of the CLIC accelerator with emphasis on the next phase is summarised in~\ref{tab:timeline_CLIC}.

\begin{table}[h!]
\small
\centering
\begin{tabular}{|l||c|}\hline
CLIC & Years \\ \hline\hline
Conceptual Design Study & 2004 -- 2012 \\
Project Implementation Plan and Readiness Report & 2013 -- 2025 \\

\hline
\hline
Project Preparation Phase 1 & 2026 -- 2028 \\
\hline
\multicolumn{2}{|c|}{Definition of the placement scenario}  \\

\multicolumn{2}{|c|}{Design optimisation and finalization}  \\

\multicolumn{2}{|c|}{Main technologies R\&D conclusions}  \\

\multicolumn{2}{|c|}{Technical Design Report -- two IPs at CERN} \\ 



\hline
\hline
Project Preparation Phase 2 & T$_0$ -- (T$_0$+5) \\
\hline 

\multicolumn{2}{|c|}{Site investigation and preparation}  \\
\multicolumn{2}{|c|}{Implementation studies with the Host states}  \\
\multicolumn{2}{|c|}{Environmental evaluation \& Project authorisation processes} \\
\multicolumn{2}{|c|}{Industrialization of key components}  \\
\multicolumn{2}{|c|}{Engineering design completion}  \\

\hline
\hline
Construction Phase (from ground breaking) & T$_1$ -- (T$_1$+10) \\
\hline
\multicolumn{2}{|c|}{Civil engineering}  \\
\multicolumn{2}{|c|}{Construction of components}  \\
\multicolumn{2}{|c|}{Installation and hardware commissioning}  \\
\hline
\hline
Beam commissioning and physics operation start & T$_1$+11 \\
\hline
\hline
\end{tabular}
 \caption{Timeline of essential development and construction steps of the CLIC project. T0 is determined by a process in 2028-29 to validate the progress and promise of the project for a further development towards implementation. T1 following Preparation Phase 2 will be determined by the processes needed, by the CERN Council and with host-states, for project approval and to start construction. The construction phase is extended by three years with respect to the technically-limited schedules to allow a transfer time into construction, and to avoid the resource conflict between HL LHC operation and initiating beam commissioning for a next collider. }
\label{tab:timeline_CLIC}
\end{table}

\subsubsection{The key technologies needed for delivery that are still under development in 2024, and the targeted performance parameters of each development}

Most of the central elements of CLIC accelerator have been developed into prototypes -- in some cases several generations of them, and tested in laboratories, beam-tests facilities, or operational machines. Overall design and performance studies have also been implemented and verified, including beam-based steering and tuning procedures.   

The challenges of the X-band technology and two beam acceleration scheme are discussed in other sections of this report, including the test and beam-facilities used to verify their performances, among others the CTF3 facility.  The nanobeam challenge encompasses several technologies and systems, from damping rings to the interaction point, from alignment and stability to instrumentation and beam dynamics. CLIC has systematically addressed all the issues and components of relevance and the status is similar for the various parts in terms of design, prototyping and beam tests. System level tests have also been implemented and successfully performed (e.g. CTF3). In many cases, for example for the damping ring systems, synchrotron sources or free electron laser linacs provide very important additional confidence and test-grounds for the performances needed. 

Many components of CLIC require little further R\&D, but require developments and further work to optimize and validate large scale industrial processes and samples, typically addressed in the pre-series phase.

\subsubsection{The critical path for technology development or design}

The most critical elements are module prototypes to verify performance and cost estimates for the X-band components at scale, and further nanobeam studies including validation of the hardware that affects such beams, e.g. alignment, stability, instrumentation, etc.   

\subsubsection{A concise assessment of the key technical risks to the delivery of the project}

The key components of the CLIC accelerator are at Technology Readiness Level (TRL) 6 ("A representative model or prototype system/subsystem is tested in a relevant environment, in our case typically a beamline or test facility, representing a major step up in a technology's demonstrated readiness") or 7 ("Prototype as part of an operational system, representing a major step up from TRL 6, requiring the demonstration of an actual system prototype in an operational environment, in our case as part of an accelerator"). A summary of the TRLs and some key improvements foreseen over the next phase are shown in~\ref{tab:CLICTRL}. 

 \begin{table}[!h]
 \centering
 \small
 \begin{tabular}{|l|c|c|}
 \hline
 \multirow{3}{*}{Component/Sub-system} & \multirow{3}{*}{TRL} & \multirow{2}{*}{Goals, risks}  \\ 
 ~&  & \multirow{2}{*}{being addressed}  \\
~ & ~ & ~  \\ \hline\hline
X-band technology & 6  & Larger systems and industrial readiness, cost and yields  \\ \hline
HOM detuning/damp  & 6 & Robustness in system tests, luminosity \\ \hline
Positron source  & 5 & Benchmarking, positron yield verification \\ \hline
Two-beam acceleration  & 7 & DB design and component developments, RF power \\ \hline
Initial emitt. and preservation  & 6 & Design, luminosity margins \\ \hline
IP spot size/stability  & 5 & Beamtests, stability and robustness    \\ \hline
 \end{tabular}
  \caption{Technology readiness and R\&D for the CLIC accelerator.}
 \label{tab:CLICTRL}
 \end{table}
 
\subsubsection{An estimate of financial and human resources needed for R\&D on key technologies}
The key elements of the CLIC accelerator activities during the Preparation Phases 1 and 2 in the timeline are summarised in~\ref{tab:CLIC2025}. The programme covers the R\&D on the critical technologies affecting the main parameters of the accelerators and technical infrastructure and/or having an impact on civil engineering, procurement of components requiring large scale production, acceptance tests and validation or production with long lead times. 

\begin{table}[!h]
\centering
\caption{Main CLIC accelerator objectives and activities in the next phase.}
\label{tab:CLIC2025}
\small
\begin{tabular}{c c}
\toprule
\textbf{Activities} & \textbf{Purpose} \\
\midrule
\multicolumn{2}{c}{\textbf{Design and parameters}}\tabularnewline  
\begin{minipage}[t]{0.46\columnwidth}
Beam dynamics studies, parameter optimisation, cost, power, system verifications in linacs and low emittance rings
\end{minipage} &
\begin{minipage}[t]{0.46\columnwidth}
Luminosity performance and reduction of risk, cost and power
\end{minipage} \tabularnewline
\\
\multicolumn{2}{c}{\textbf{Main linac modules}}\tabularnewline 
\begin{minipage}[t]{0.46\columnwidth}
Construction of $\sim5$ prototype modules in qualified industries, optimised design of the modules with their supporting infrastructure in the main linac tunnel
\end{minipage} & %
\begin{minipage}[t]{0.46\columnwidth}
Final technical design, qualification of industrial partners, production models, performance verification 
\end{minipage} \tabularnewline
\\
\multicolumn{2}{c}{\textbf{Accelerating structures}}\tabularnewline
\begin{minipage}[t]{0.46\columnwidth}
Production of $\sim50$ accelerating structures, including structures for the modules above
\end{minipage} & %
\begin{minipage}[t]{0.46\columnwidth}
Industrialisation, manufacturing and cost optimisation, conditioning studies in test-stands  
\end{minipage} \tabularnewline
\\
\multicolumn{2}{c}{\textbf{Operating X-band test-stands, high efficiency RF studies}}\tabularnewline  
\begin{minipage}[t]{0.46\columnwidth}
Operation of X-band RF test-stands at CERN and in collaborating institutes for structure and component optimisation, further development of cost-optimised high efficiency klystrons. 
\end{minipage} & %
\begin{minipage}[t]{0.46\columnwidth}
Building experience and capacity for X-band components and structure testing, validation and optimisation of these components, cost reduction and increased industrial availability of high efficiency RF units 
\end{minipage} \tabularnewline
\\
\multicolumn{2}{c}{\textbf{Other technical components}}\tabularnewline
\begin{minipage}[t]{0.46\columnwidth}
Magnets, instrumentation, alignment, stability, vacuum    
\end{minipage} &
\begin{minipage}[t]{0.46\columnwidth}
Luminosity performance, costs and power, industrialisation 
\end{minipage} \tabularnewline
\\
\multicolumn{2}{c}{\textbf{Drive beam complex and main beam injector studies, incl. L\&S band RF sources}}\tabularnewline  
\begin{minipage}[t]{0.46\columnwidth}
Drive-beam detailed design and component prototypes for drivebeam and main beam injectors. 
\end{minipage} &
\begin{minipage}[t]{0.46\columnwidth}
Verification of the most critical parts of the drive-beam and injectors, among them further development of industrial capabilities for high efficiency L-band klystrons
\end{minipage} \tabularnewline
\\
\multicolumn{2}{c}{\textbf{Civil Engineering, siting, infrastructure}}\tabularnewline  
\begin{minipage}[t]{0.46\columnwidth}
Detailed site specific technical designs, site preparation, environmental impact study and corresponding procedures in preparation for construction
\end{minipage} &
\begin{minipage}[t]{0.46\columnwidth}
Preparation for civil engineering works, obtaining all needed permits, preparation of technical documentation, tenders and commercial documents
\end{minipage} \tabularnewline
\\
\bottomrule
\end{tabular}
\end{table}

Civil engineering and infrastructure preparation will become increasingly detailed during the preparation phase. The first step is final placement of the collider complex and its infrastructure after assessment of their territorial compatibility, both for its initial phase and potential upgrades. This step is particularly urgent to establish that a linear collider such as CLIC can be constructed at CERN as outlined in this report.  This will be followed by optimisation of the placement scenario with the relevant regional and local stakeholders as a part of wider implementation studies with the Host Regions/States. An environmental impact study and corresponding public inquiry will be needed as a prerequisite to authorisations for construction. 



\begin{figure}[!h]
\centering
\includegraphics[width=0.7\textwidth]{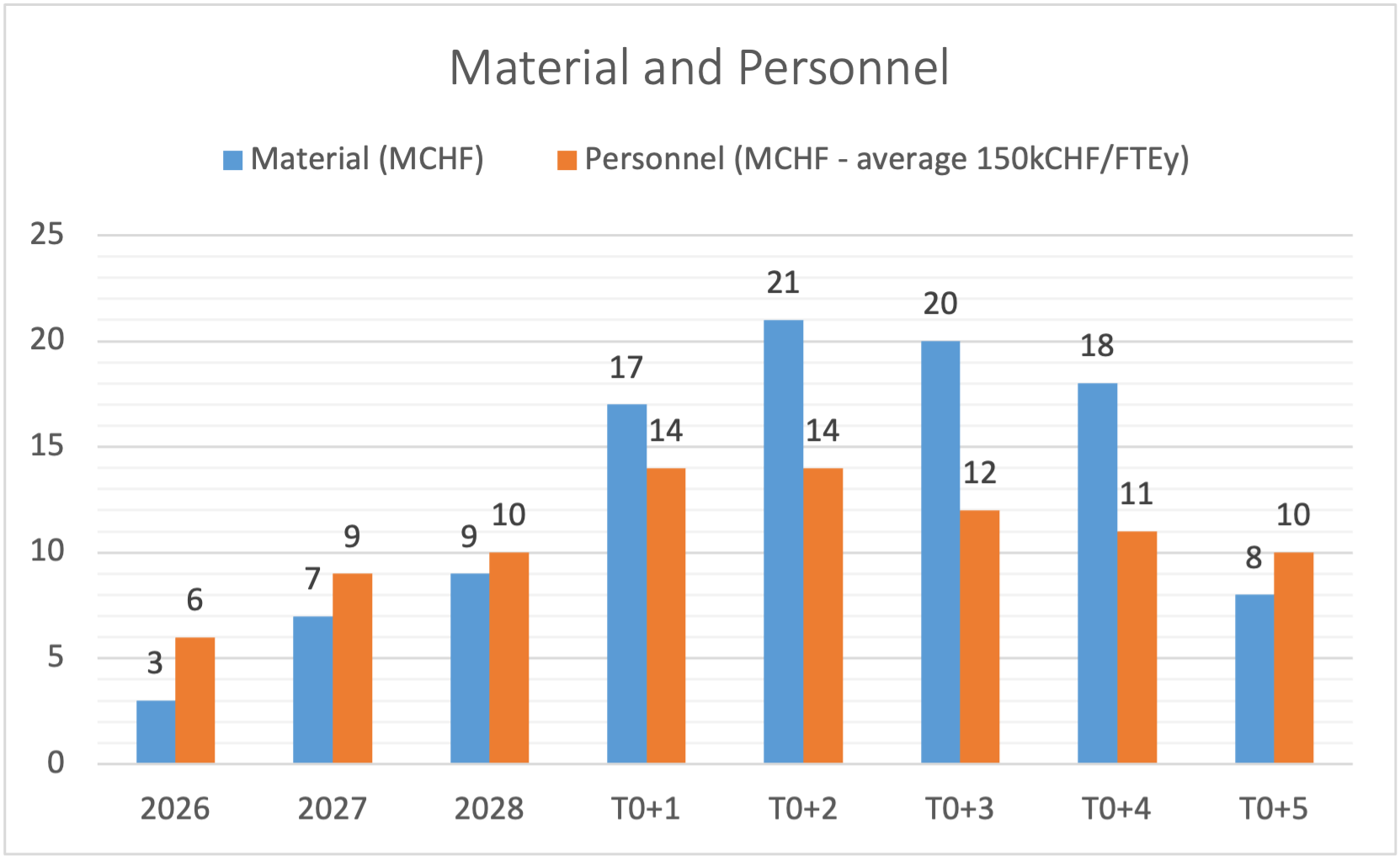}
\caption{Resource timeline for accelerator R\&D in the next phases (Phase 1 and 2 in the timeline) addressing the studies described in ~\ref{tab:CLIC2025}. The use of construction resources after a potential approval is not included.}
\label{fig_R&Dres}
\end{figure}

The resources for such a programme are summarised in~\ref{fig_R&Dres}, showing that the initial part of the programme (Phase 1 in the time-line) requires limited resources while the larger industrial prototyping and pre-series efforts occur later (Phase 2), combined with construction preparation. Around 1/3 of these resources could be available outside CERN. 

In parallel, civil engineering preparation to be able to place time-critical contracts to start the construction, including continued environmental studies, will require significant resources during the second parts of Phase 2, typically 5\% of the civil engineering budget.

\subsection{Dependencies}

\subsubsection{Whether a specific host site is foreseen, or whether options are available}


CLIC is studied for implementation at CERN. As it is independent of the existing accelerator complex, it can also be implemented in other places, though likely at a sizeable penalty in time and cost (see next bullet point).

\subsubsection{The dependencies on existing or required infrastructure}

Although CLIC will not use existing CERN accelerators as injectors, its implementation relies heavily on CERN's infrastructure, expertise, and institutional framework built over 70 years. The CERN Convention provides a solid legal and organisational foundation for international collaboration, funding, and project governance.

CERN's technical facilities, test beams, and experience in prototyping, quality control, and industrial coordination are essential for CLIC component development. The organisation's familiarity with local regulations, environmental and safety standards, and labour frameworks are crucially important for implementing the project.

Very important for CLIC is that CERN owns suitable land, with access to key resources such as electrical power, water, communication networks, and transport infrastructure, making it a very well adapted host site.

\subsubsection{The technical effects of project execution on the operations of existing infrastructures at the host site}
The CLIC accelerator complex at CERN is fully independent of the existing CERN accelerator complex and associated infrastructure. The layout and work at the Pr\'{e}vessin site must be carefully planned to avoid and minimize temporary construction-related conflicts. 

\subsection{Commentary on Current Project Status}
\label{addendum:status}

\subsubsection{A concise description of the current design / R\&D / simulation activities leading to the project, and the community pursuing these}

The CLIC collaboration~\cite{clic-study} focuses on advancing X-band technology, with ongoing improvements to RF networks and high-efficiency power units. Compact linac designs based on CLIC technology are being studied for applications as light and neutron sources, supporting broader technology deployment and industrial readiness of key components.

Work on luminosity performance covers the 380\,GeV stage and higher energies, with progress in damping ring and beam delivery system design, emittance preservation, and positron production. Power optimisation efforts target both system design and hardware development, profiting from the advancement of high-efficiency klystrons.

Lifecycle assessments (LCA) to evaluate civil engineering requirements, operational energy use, and environmental impacts, contributing to a comprehensive understanding of the sustainability of CLIC.

The long back-up document, the CLIC Readiness Review~\cite{Adli:ESU25RDR}, contains a much more detailed description of progress on these points and more. 

R\&D programs for detector technologies meeting the CLICdet requirements are performed within the CLICdp collaboration~\cite{clicdp} and other international collaborative frameworks, in line with the 2021 ECFA Detector Roadmap. Recent progress on Detector R\&D for CLIC is reported in~\cite{CLIC-det-progress:2025}.

\subsubsection{A statement of any major in-kind deliverables already negotiated}

At this stage no major in-kind deliverables are negotiated for construction. Inside the collaboration there are key collaboration partners with technical skills and on-going related activities that are candidates for the various deliverables. 

\subsubsection{Any other key technical information points in addition to those captured above, including references to additional public documents addressing the points above.}

The long back-up document, the CLIC Readiness Review (updated version is available at~\cite{Adli:ESU25RDR}), contains key technical information. 

A list of CLIC overview documents, including previous ESPPU submissions can be found at the following location: 

\vspace*{-3mm}
\begin{center}
\textbf{Overview of main CLIC documents: \url{http://clic.cern/european-strategy}} 
\end{center}


\FloatBarrier
\subsection{Community}\label{addendum:community}

The CLIC accelerator collaboration and CLIC Detector and Physics collaboration together
comprise around 400 participants from approximately 75 institutes worldwide~\cite{clic-study,clicdp}.
Additional contributions are made from beyond the collaborations. 

Editor list for this document: 
Erik Adli (University of Oslo and CERN),  
Gerardo D'Auria (Sincrotrone Trieste),
Nuria Catalan Lasheras (CERN),  
Vera Cilento (John Adams Institute for Accelerator Science, University of Oxford and CERN),  
Roberto Corsini (CERN),  
Dominik Dannheim (CERN),  
Steffen Doebert (CERN),  
Mick Draper (CERN),  
Angeles Faus-Golfe (IJCLab, Orsay),
Edward Fraser Mactavish (CERN),  
Alexej Grudiev (CERN),  
Andrea Latina (CERN),  
Lucie Linssen (CERN),  
John Andrew Osborne (CERN),  
Yannis Papaphilippou (CERN),  
Philipp Roloff (CERN),  
Aidan Robson (Univ. Glasgow),  
Carlo Rossi (CERN),  
Andre Sailer (CERN),  
Daniel Schulte (CERN),  
Eva Sicking (CERN),  
Steinar Stapnes (CERN),  
Igor Syratchev (CERN),  
Rogelio Tomas Garcia (CERN),  
Walter Wuensch (CERN).

A complete author list for the CLIC Readiness Report is under preparation.



\end{document}